\documentclass[12pt]{article}
\usepackage[T1]{fontenc}
\usepackage[latin9]{inputenc}
\usepackage{amsthm, latexsym, epsfig, verbatim, amsmath, amssymb, lscape,graphics,booktabs,multirow,array,threeparttable,ragged2e,color}
\usepackage{natbib}
\usepackage{titling}
\usepackage{titlefoot}
\usepackage{sectsty}
\usepackage{authblk}
\usepackage{enumitem}

\usepackage{caption}
\allowdisplaybreaks

\sectionfont{\fontsize{13}{13}\selectfont\centering}
\renewcommand\thesection{\arabic{section}.}
\subsectionfont{\fontsize{12}{12}\selectfont}
\renewcommand\thesubsection{\thesection\arabic{subsection}.}

 \theoremstyle{remark}
 \newtheorem{rem}{\protect\remarkname}
\theoremstyle{plain}
\newtheorem{thm}{\protect\theoremname}
 \theoremstyle{definition}
 
 \theoremstyle{plain}
 
 \theoremstyle{assumption}
 \newtheorem{assum}{\protect\assumptionname}
\setlist[description]{font=\normalfont\space}

\usepackage{booktabs,calc}

\providecommand{\examplename}{Example}
 \providecommand{\lemmaname}{Lemma}
 \providecommand{\remarkname}{Remark}

\providecommand{\theoremname}{Theorem}
\providecommand{\assumptionname}{Assumption}
\newtheorem{cond}{Condition}

\newcommand{\bv}{{\boldsymbol v}}
\newcommand{\bW}{{\boldsymbol W}}
\newcommand{\bx}{{\boldsymbol x}}
\newcommand{\bX}{{\boldsymbol X}}

\newcommand{\balpha}{{\boldsymbol \alpha}}
\newcommand{\bbeta}{{\boldsymbol \beta}}
\newcommand{\bfeta}{{\boldsymbol \eta}}
\newcommand{\bgamma}{{\boldsymbol \gamma}}

\newcommand{\btheta}{{\boldsymbol \theta}}
\newcommand{\bzero}{{\boldsymbol 0}}

\newcommand{\wlambda}{\widehat\lambda}
\newcommand{\wLambda}{\widehat\Lambda}
\newcommand{\wbalpha}{\widehat{\boldsymbol \alpha}}
\newcommand{\wbgamma}{{\widehat\bgamma}}
\newcommand{\wbtheta}{{\widehat\btheta}}
\newcommand{\tL}{{\widetilde L}}

\newcommand{\tbX}{{\widetilde{\boldsymbol X}}}
\newcommand{\tbx}{{\widetilde{\boldsymbol x}}}
\newcommand{\calA}{{\cal A}}

\newcommand{\calO}{{\cal O}}
\newcommand{\wcalA}{{\widehat\calA}}

\newcommand{\mathR}{{\mathbb R}}

\usepackage{prettyref}
\newrefformat{tab}{Table \ref{#1}}
\newrefformat{fig}{Figure \ref{#1}}
\newrefformat{thm}{Theorem \ref{#1}}
\newrefformat{lem}{Lemma \ref{#1}}

\setcitestyle{aysep={},yysep={;}}

\makeatletter
\def\@seccntformat#1{\@ifundefined{#1@cntformat}%
 {\csname the#1\endcsname\quad} % default
 {\csname #1@cntformat\endcsname}% enable individual control
}
\let\oldappendix\appendix %% save current definition of \appendix
\renewcommand{\appendixname}{APPENDIX}
\renewcommand\appendix{%
 \oldappendix
 \newcommand{\section@cntformat}{\appendixname~\thesection\quad}
}
\makeatother

\begin{document}
\def\spacingset#1{\renewcommand{\baselinestretch}%
{#1}\small\normalsize} \spacingset{1}

\title{\bf{\large Defining and Estimating Subgroup Mediation Effects with Semi-Competing Risks Data}}
\author{\normalsize Fei Gao} \affil{\normalsize Fred Hutchinson Cancer Research Center, Seattle, WA, USA.} 
\author{\normalsize Fan Xia and Kwun Chuen Gary Chan}
\affil{\normalsize Department of Biostatistics, University of Washington, Seattle, WA, USA.}
%\affil{Department of Biostatistics, University of Washington, Seattle, WA 98195}
\date{}
\bigskip

\begin{titlepage}

\maketitle

\begin{abstract}
In many medical studies, an ultimate failure event such as death is likely to be affected by the occurrence and timing of other intermediate clinical events.
Both event times are subject to censoring by loss-to-follow-up but the nonterminal event may further be censored by the occurrence of the primary outcome, but not vice versa.
To study the effect of an intervention on both events, the intermediate event may be viewed as a mediator, but conventional definition of direct and indirect effects is not applicable due to semi-competing risks data structure.
We define three principal strata based on whether the potential intermediate event occurs before the potential failure event, which allow proper definition of direct and indirect effects in one stratum whereas total effects are defined for all strata.
We discuss the identification conditions for stratum-specific effects, and proposed a semiparametric estimator based on a multivariate logistic stratum membership model and within-stratum proportional hazards models for the event times.
By treating the unobserved stratum membership as a latent variable, we propose an EM algorithm for computation.
We study the asymptotic properties of the estimators by the modern empirical process theory and examine the performance of the estimators in numerical studies. 

\end{abstract}

\noindent%
\vfill

%\unmarkedfntext{Fei Gao is Assistant Member at Vaccine and Infectious Disease Division, Fred Hutchinson Cancer Research Center, Seattle, WA 98109(E-mail: \emph{fgao@fredhutch.org}),
%Fan Xia is Ph.D. student (E-mail: \emph{fanxia@uw.edu}), Gary KC Chan is Professor (E-mail: \emph{kcgchan@uw.edu}), Department of Biostatistics, University of Washington, Seattle, WA 98195.
%This work was supported by the National Institutes of Health awards R01GM047845, R01AI029168, R01CA082659, and P01CA142538.
%}
%
\end{titlepage}

\newpage
\clearpage
\setcounter{page}{1}
\spacingset{1.55} % DON'T change the spacing!

\section{INTRODUCTION}	
Evaluating the causal effects of an intervention on a clinical outcome is a common theme in many medical studies. After an overall relationship between the intervention and outcome is established, it is often of further interest to understand the biological or mechanistic pathways that contribute to the causal treatment effect.
Causal mediation analysis is often utilized to disentangle the total treatment effect by decomposing it into the indirect effect, i.e., the effect exerted by intermediate variables (mediators), and the direct effect, i.e., the effect involving pathways independent of the hypothesized mediators.
A number of methods were proposed for causal mediation analysis with survival outcomes, for a single mediator measured at study entry \citep{lange2011direct, vanderweele2011causal, tchetgen2011causal, lange2012simple} and for longitudinal mediators
\citep{lin2017mediation, zheng2017longitudinal,  didelez2019defining, vansteelandt2019mediation, aalen2020time}.
%longitudinal mediators under stochastic intervention or separable intervention components different from traditional definition of causal mediation effects.

In many biomedical studies, intermediate, non-terminal landmark events are recorded in addition to the primary failure event because they are important to evaluate prognosis.
Due to the ordering of the two events, the non-terminal event is subject to censoring by the occurrence of the terminal event, but not vice versa, such that semi-competing risks data are observed \citep{fine2001semi}.
In this paper, we consider a setting where a non-terminal event may serve as a mediator for individuals to whom the event would occur before the terminal event.
An example is a multi-center trial of allogeneic bone marrow transplants in patients with acute leukemia \citep{copelan1991treatment, klein2006survival}, where the primary interest is on the effect of different treatment regimen (methotrexate + cyclosporine vs methylprednisolone + cyclosporine) on the survival time.
The event time of an intermediate endpoint, chronic graft-versus-host disease (GVHD), is a major side effect of the transplant that can be lethal.
However, some patients died without experiencing GVHD, such that GVHD event time is subject to censoring by the death time.

Causal mediation analysis with semi-competing risks data is particularly challenging.
First, the mediator is only well-defined for those who would have the non-terminal event developed before the occurrence of the primary event.
Therefore, the conventional definition of natural indirect and direct effects based on replacing the counterfactual of mediator under one treatment by that under the other can hardly apply to the entire population.
This challenge is similar to that of the `truncation-by-death' problem \citep{zhang2003estimation,comment2019survivor}, where the primary outcome is only available if the terminal event does not occur.
However, there is a substantial difference in that the primary outcome of interest is the terminal event in our setting.
%For instance, \cite{comment2019survivor} considered estimation of the treatment effect on the intermediate variable under `truncation-by-death' of the ultimate outcome.
Moreover, the semi-competing risks data structure, that is, the primary event may censor the intermediate event but not vice versa, posts additional challenges in the identifiability of parameters. %ince the primary event may censor the intermediate event but not vice versa, the identifiability of parameters is challenging as we shall discuss later. 
%the joint distribution of the two event times is only identifiable on the upper wedge of the support.
%Therefore, the identifiability of certain functionals involved in the construction of the mediation effects needs special attention.
Upon finishing this paper, we became aware of the newly accepted paper by \cite{huang2020causal} which considered this problem by a counting process framework.
The problem formulation, estimand and assumptions are all very different from our work.
For instance, we do not make sequential ignorability assumptions on surviving subpopulations at arbitrary post-treatment time, because those evolving subpopulations are generally healthier than the baseline study population before the treatment is assigned.

In this paper, we consider a novel principal stratification approach to define causal mediation effects in the subgroup where the intermediate event happens when given either treatment, i.e., those susceptible to the intermediate event under both treatments. The notations and settings are given in Sections 2.1 and 2.2.
We discuss the identification conditions needed for estimating the stratum-specific natural indirect and direct effects in Section 2.3, and proposed a semiparametric estimator based on a multivariate logistic stratum membership model and within-stratum proportional hazards models for the event times in Section 2.4.
By treating the unobserved stratum membership as a latent variable, we propose an EM algorithm for computation of the nonparametric maximum likelihood estimator in Section 2.5.
We also study the asymptotic properties of the estimators using the modern empirical process theory in Section 2.6 and examine the performance of the estimators in simulation studies Section 3.  An analysis of data from a clinical trial is given in Section 4, and concluding remarks are included in Section 5.  Proofs and detailed derivations are given in the Appendix.

\section{Methods}
\subsection{Notations for Observed Data}
Let $A$ be a binary treatment,  $T$ be the time to a primary event of interest and $M$ be the time to an intermediate, non-terminal event.
%The treatment effect of $A$ on $T$ may be mediated by an intermediate event, whose time is denoted by $M$.
The intermediate event time $M$ may be censored by the occurrence of the primary event, but not vice versa, such that we observe semi-competing risks data.
%In the semi-competing risks data literature, the primary event is called a terminal event, and the secondary intermediate event is called a nonterminal event.
For example, $A$ is a treatment for prolonging survival time, $T$ is the time to death, and $M$ is the time to cancer progression.
The occurrence of death may censor the cancer progression onset, but not vice versa.

Let $\bX$ be a collection of baseline covariates that may be associated with either or both events.
Let $C$ denote a censoring time for the primary event, for example, end of follow-up time.
Then, we observe $Y \equiv \min(T, C)$ and $\Delta^T = I(T\le C)$ for the primary event, and $Z \equiv\min(M, Y)$ and $\Delta^M = I(M\le Y)$ for the intermediate event.
The observations are versions of the counterfactual variables that we define as follows.

\subsection{Counterfactuals and Causal Estimands} \label{sec:strata}
%Our primary goal is to investigate how the treatment $A$ and the nonterminal event time $M$ influence the terminal event time $T$ after adjusting for baseline covariates $\bX$, through mediation modeling.
To define causal mediation effects of interest, we adopt the potential outcomes framework.
In conventional causal mediation analysis based on counterfactuals, $M(a)$ denotes the counterfactual nonterminal event time when the treatment is set to $a$ and $T(a, m)$ denotes the counterfactual  terminal event time when the treatment is set to $a$ and the nonterminal event time (mediator) is set to $m$.
A comparison of $T(a, M(a))$ with $T(a, M(a^*))$ would define a measure of the natural indirect effect of changing the mediator from $M(a)$ to $M(a^*)$ and a comparison of $T(a, M(a^*))$ with $T(a^*, M(a^*))$ would define a measure of the natural direct effect of changing the treatment from $a$ to $a^*$.
Both natural indirect and direct effects involve the term $T(a, M(a^*))$, i.e., the counterfactual outcome for the terminal event time when the treatment is set to $a$ and the nonterminal event time is set to $M(a^*)$, the counterfactual nonterminal event time when the treatment is set to $a^*$.  However, these conventional definitions are inadequate for semi-competing risk settings and needs to be modified for the following reasons.
When the potential primary event happens before the potential intermediate event, the value of the mediator is not well-defined (and is often set to $\infty$ by convention) and in such a case the potential primary event time shall not be dependent on an arbitrary $m$ greater than the potential primary event time.
Furthermore, although the single-world variables $M(a)$ (which may be $\infty$) and $T(a)=T(a,M(a))$, $a=0,1$, are well-defined for any subjects in general, $T(a,M(a^*))$ requires additional cross-world assumptions to be always well-defined for every subjects.
In light of these reasons, we consider the following cross-world ordering invariance assumption:

\begin{assum}%[Cross-world ordering invariance]
For any $a, a^* \in \{0,1\}$, either
(i) $M(a)<\infty$ and $T(a^*,M(a))\ge M(a)$ or
(ii) $M(a)=\infty$ and $T(a^*,M(a))<M(a)$ holds.
\end{assum}

Moreover, the potential non-terminal event may or may not occur before the potential primary event time under different treatment assignments.
Due to these considerations, we examine causal effects based on our proposed principal stratification approach,  extended from \cite{frangakis2002principal}.
Intuitively, we stratify the study population into latent classes identified by $U$ with 3 categories based on whether they are susceptible to the nonterminal event under different treatment assignments:
\begin{enumerate}
\item $U=1$: $M(0)\le T(0)$ and $M(1)\le T(1)$ (always susceptible).
\item $U=2$: $M(0)\le T(0)$ and $M(1)=\infty$ (prevented).
\item $U=3$: $M(0)=\infty$ and $M(1)=\infty$ (always non-susceptible).
\end{enumerate}
Here, we do not have a fourth stratum: $M(0)=\infty$ and $M(1)\le T(1)$, such that the treatment never convert a subject from non-susceptible to susceptible.
This restriction is along the same line as the ``no defier'' assumption commonly adopted in the instrumental variables methods, suggesting that the treatment effect is ``monotone'' and no reversed effect for the subjects \citep{angrist1996identification}.
\begin{rem}
The defined strata (and associated stratum-specific effects) are substantially different from the survivors' principal stratum (and the survivor average causal effect (SACE)) that is commonly defined in ``truncation by death'' literature \citep{zhang2003estimation, comment2019survivor}.
In particular, the survivors' principal stratum is defined by $\{T(0)\ge t, T(1)\ge t\}$ for some fixed time $t$ in \cite{comment2019survivor}, whereas our definition does not depend on any arbitrary post-treatment time $t$.
\end{rem}

\begin{rem}
\cite{lin2017mediation} explained the difficulties in defining natural mediation effects in survival context with longitudinal mediators.  They defined interventional effects in a discrete-time setting, where the mediators and past survival status are subject to a hypothetical intervention.  They mentioned that principal stratification as an alternative framework to avoid such hypothetical intervention, but did not explore further.   We consider a different setting that shares some of the difficulties, but also with unique data structure so that principal strata can be defined.
\end{rem}
Under suitable assumptions (to be made clear in Section 2.3), for $U=1$, the joint distribution of $(T(a),M(a)), a=0,1$ could be nonparametrically identified on the upper wedge of the positive quadrant, and by cross-world invariance, $T(1,M(0))\ge M(0)$ is well-defined in the same region, and therefore we can define and estimate the stratum-specific natural indirect and direct effects: 
%In light of these observations, for stratum with $U=1$, we define the stratum-specific natural indirect and direct effects as
\begin{equation}
NIE_1(t;\bx) = \Pr \{T(1,M(1))\ge t | \bX=\bx, U=1\} - \Pr \{T(1,M(0))\ge t | \bX=\bx, U=1\}\label{equ:NIE1}
\end{equation}
and
\begin{equation}
NDE_1(t;\bx) =\Pr \{T(1,M(0))\ge t | \bX=\bx, U=1\} - \Pr \{T(0,M(0))\ge t | \bX=\bx, U=1\}.\label{equ:NDE1}
\end{equation}
In stratum with $U=2$, although the pair $(T(1,M(0)),M(0))$ is technically defined on the upper wedge, the joint distribution of $(T(1),M(1))$ is not identified in that region as $M(1)=\infty$, we therefore do not seek to estimate the indirect and direct effects, but
%is not well-defined because $T(1,m)$ is only defined for $m=\infty$ and $M(0)<\infty$ with probability 1.  However, $T(a)=T(a,M(a))$ is still well-defined and 
the stratum-specific total effect can still be estimated:
\[TE_2(t;\bx) = \Pr \{T(1) \ge t | \bX = \bx, U = 2\} - \Pr \{T(0) \ge t | \bX = \bx, U = 2\} \ .\]
In stratum with $U=3$, $M(0)=M(1)=\infty$ and $T(1,M(0))=T(1,M(1))$ so there is no indirect effect.  The stratum-specific total effect can still be defined as
\[TE_3(t;\bx) = \Pr \{T(1) \ge t | \bX = \bx, U = 3\} - \Pr \{T(0) \ge t | \bX = \bx, U = 3\}.\]
\begin{rem}
In principle, a mediator shall satisfy temporal precedence, that it shall occur before the primary event.  Therefore, one can view that the mediator is technically absent in $U=3$, and an attempt to define mediation effects would be futile.  In $U=2$, the presence of the mediator before the primary event only happens in one treatment level with certainty. As a result, one cannot fix the mediator level at a different treatment level, and mediation effects cannot be defined.
Note that in $U=2$, $TE_2$ can be interpreted as the treatment effect in survival among individuals whose mediating events are prevented by the treatment.
\end{rem}
\subsection{Identification}
To identify the stratum-specific natural indirect and direct effects and stratum-specific total effects, we impose the following assumptions.
\begin{assum}%[Consistency]
\label{ass:consistent}
If $A = a$, then $M = M(a)$ and $T = T(a)$ with probability one.
\end{assum}
%\begin{assum}[Conditional Exchangeability] \label{ass:exchange}
%For $a = 0, 1$, 
%\[\{M(a), T(a)\}\perp A | \bX.\]
%\end{assum}
\begin{assum}%[Sequential Ignorability within Stratum]
\label{ass:seq_ign}
For $a, a^* \in \{0,1\}$ and $u\in \{1,2,3\}$, \begin{equation}
\{T(a, M(a^*)), M(a^*)\}\perp A| \bX, U=u \label{equ:seq1}
\end{equation}
and
\begin{equation}
\Pr(T(a,M(a^*))| M(a^*)=m,A=a,\bX,U=u) = \Pr(T(a,M(a))| M(a)=m,A=a,\bX,U=u).\label{equ:seq2}
\end{equation}
\end{assum}
Assumptions \ref{ass:consistent} is the standard consistency assumption for causal inference.
Assumption \ref{ass:seq_ign} serves a similar purpose as the sequential ignorability assumption \citep{imai2010identification} but is weaker so that the the assumption holds within a stratum and only requires $T(a,M(a^*))$ to be well-defined.
%For $U=1$, Assumption \ref{ass:seq_ign} implies the sequential ignorability assumption commonly employed in mediation analysis.
%For $U=2,3$, Assumption \ref{ass:seq_ign} reduces to the standard assumption of conditional exchangeability within stratum.
Based on Assumptions \ref{ass:consistent} - \ref{ass:seq_ign}, we are able to connect the stratum-specific natural indirect and direct effects and stratum-specific total effects with the distribution of the observed data given stratum membership as follows.
%The results are stated in the following Theorem.
\begin{thm}\label{thm:express}
Under Assumptions \ref{ass:consistent} - \ref{ass:seq_ign}, for stratum with $U=1$, the stratum-specific natural indirect effect $NIE_1(t;\bx)$ is equal to
\begin{align*}
& \int_0^t \left\{1-\Pr(T< t|M = m, \bX=\bx, A=1, U=1)\right\}\\
&\qquad \times \left\{d F_{M|\bX=\bx, A=1, U=1}(m) - d F_{M| \bX=\bx, A=0, U=1}(m)\right\}\\
&\qquad + \Pr(M\le t|\bX=\bx, A=0, U=1)-\Pr(M\le t|\bX=\bx, A=1, U=1),
\end{align*}
and the stratum-specific natural direct effect $NDE_1(t;\bx)$ is equal to
\begin{align*}
& \int_0^t \left\{\Pr(T< t| M= m, \bX=\bx, A=0,U=1) \right.\\
&\qquad \left.- \Pr(T< t| M = m, \bX=\bx, A=1, U=1)\right\}d F_{M|\bX=\bx, A=0,U=1}(m).
\end{align*}
Under Assumptions \ref{ass:consistent} and \ref{ass:seq_ign}, for stratum with $U=2$, the stratum-specific total effect $TE_2(t;\bx)$ is equal to
\[\Pr(T\ge t| A=1,\bX=\bx,U=2) - \Pr(T\ge t| A = 0,\bX=\bx,U=2);\]
and for stratum with $U=3$, the stratum-specific total effect $TE_3(t;\bx)$ is equal to
\[\Pr(T\ge t| A=1,\bX=\bx,U=3) - \Pr(T\ge t| A = 0,\bX=\bx,U=3).\]
\end{thm}
The proof of Theorem \ref{thm:express} is given in Appendix \ref{append:express}.
Since $U$ is unobserved, we cannot use Theorem 1 directly to identify those stratum-specific effects from observed data.  To do so, one would further assume
%Based on the expressions in Theorem \ref{thm:express}, we are able to identify those stratum-specific effects if some model restrictions are imposed to the stratum-specific distributions on the observed data.
%Particularly, one would assume
\begin{assum}\label{ass:mem}
With probability one, $U$ is conditional independent of $A$ given $\bX$.
\end{assum}
\begin{assum}
\label{ass:model} With probability one, 
\begin{align*}
 \Pr (M(0) = m|\bX = \bx,U=2) =& g_1\left\{\Pr (M(0)= m|\bX = \bx,U=1);\bx\right\},\\
 \Pr (T(0)\ge t|M(0)=m,\bX = \bx,U=2) =& g_2\left\{\Pr (T(0)\ge t|M(0) = m,\bX = \bx,U=1);\bx\right\},\\
 \Pr (T(1)\ge t|\bX = \bx,U=2) =& g_3\left\{\Pr (T(1)\ge t|\bX = \bx,U=3);\bx\right\},
\end{align*}
for some known functions $g_k(\cdot;\bx)$ $(k=1,2,3)$.
\end{assum}

\begin{assum} \label{ass:cond_ind_cens}
$(M, T, U)$ is conditionally independent of $C$ given $A$ and $\bX$, and the upper bound of the support of $T$ is no larger than that of $C$.
%For $a=0, 1$, $\{M(a), T(a)\}\perp C | \bX$.
%Given $U=1$, for $a=0,1$, the upper bound of the support of $T(a)$ is no larger than that of $C$.
%Given $U=2$, the upper bound of the support of $T(0)$ is no larger than that of $C$.
\end{assum}

Assumption \ref{ass:mem} requires that the stratum membership is not affected by the treatment assignment $A$ given covariates $\bX$.
Assumption \ref{ass:model} requires some knowledge on the relationships of stratum-specific event time distributions.
The first part of Assumption \ref{ass:cond_ind_cens} is a standard assumption for non-informative censoring time.
The second part of Assumption \ref{ass:cond_ind_cens} is an extension of the independent censoring and sufficient follow-up assumption in \cite{maller1992estimating} for nonparametric estimation of cured proportion in censored data.
The assumption on the upper bounds of the supports ensures sufficient observation of the tail behaviour of the event times for identification of stratum membership.
%Even though the stratum-specific conditional independence assumption precludes additional correlation of the event times (e.g. usage of frailty models), it is sensible since additional correlation cannot be identified based on semi-competing risks data with unobserved stratum membership.
By further assuming Assumptions \ref{ass:mem} - \ref{ass:cond_ind_cens}, we obtain the identification results in Theorem \ref{thm:ident}, whose proof is given in Appendix A.2.
\begin{thm}\label{thm:ident}
Under Assumptions \ref{ass:consistent} - \ref{ass:cond_ind_cens}, the stratum-specific effects can be identified, with identification formulas given in Appendix A.2.
\end{thm}

%\begin{remark}
%For subjects with $U=1$, the support of the outcomes $(M,T)$ is the upper wedge $\{(m,t):0< m\le t<\infty\}$.
%Because of right-censoring, the observed semi-competing data provide information only on the bounded upper wedge $\{(m,t):0<m\le t\le \tau\}$, where $\tau$ is the maximum follow-up period.
%The resulting expressions in Theorem \ref{thm:express} involve only quantities on the identifiable region, such that the stratum-specific natural indirect and direct effects for stratum with $U=1$ can be identified based on observed semi-competing risks data.
%\end{remark}

Theorem \ref{thm:ident} gives the identification result based on nonparametric models for $U$ and for $(M,T)$ given $U$ with minimum assumptions.
In particular, Assumption 4 requires some modeling assumptions to be made.
In practice, we may consider additional model assumptions for $U$ and $(M,T)$ to gain power in understanding the causal effects.
In the next section, we extend the multistate modeling idea in the literature of semi-competing risks data to form such a model.

\subsection{Modeling assumptions}\label{sec:PHmodel}
One way to model semi-competing risks data is to use a multistate framework \citep{xu2010statistical}.
In multistate analysis of semi-competing risks data, usually three states (states 1 - 3) are involved, corresponding to healthy (state 1), illness (state 2), and death (state 3) in an illness-death model.
All subjects starts at state 1. A subject enters state 2 if he/she develops the intermediate event, while he/she enters state 3 if he/she develops the primary event.
In traditional illness-death model for semi-competing risks data, three processes moving from one state to another are modeled: (1) healthy to illness (state 1 to 2), (2) illness to death (state 2 to 3), and (3) healthy to death (state 1 to 3).

Here, we extend the idea and model the processes moving from one state to another in different strata defined in Section \ref{sec:strata}.
For subjects with $U=1$ and subjects with $U=2$ receiving $A=0$, the processes of healthy to illness and illness to death are involved and we model the time to the nonterminal event $M$ and the residual time $R\equiv T-M$.  We assume that $M$ and $R$ are conditionally independent given $A, \bX$, and $U$.  This serves two purposes: to obtain a tractable EM algorithm in Appendix B, and to avoid the problem of induced informative censoring caused by residual dependence between $M$ and $R$ \citep{wang1998nonparametric, lin1999nonparametric}.
For subjects with $U=2$ receiving $A=1$ and subjects with $U=3$, the process of healthy to death is involved.
This proposed model is related to but different from the illness-death model, in that the transition structure depends on the principal strata in our proposed model.

Suppose that for a subject with $U=1$, the nonterminal event time follows a proportional hazards model with hazard function given by
\[\lambda_M^{(1)}(t|A=a, \bX=\bx) =\lambda_1(t)\exp\left(\beta_{M1}a + \bgamma_{M1}^{\rm T}\bx\right), \]
and the gap time between the occurrences of nonterminal and terminal events $R$ follows a proportional hazards model with hazard function given by
\[\lambda_R^{(1)}(r|A=a, \bX=\bx) =\lambda_2(r)\exp\left(\beta_{R1}a + \bgamma_{R1}^{\rm T}\bx \right).\]
Suppose that for subject with $U=2$ and unexposed to treatment $(A=0)$, the nonterminal event time follows a proportional hazards model with hazard function given by
\[\lambda_M^{(2)}(t|A=0, \bX= \bx) =\lambda_1(t)\exp\left(\beta_{M2} + \bgamma_{M2}^{\rm T}\bx\right), \]
and the gap time between the occurrences of nonterminal and terminal events follows another proportional hazards model with hazard function given by
\[\lambda_R^{(2)}(r|A=0, \bX=\bx) =\lambda_2(r)\exp\left(\beta_{R2} + \bgamma_{R2}^{\rm T}\bx\right).\]
Here, subjects with $U=1$ and subjects with $U=2$ unexposed to treatment share the same baseline hazard functions, although the hazard ratios for covariates may be different.
The parameters $\beta_{M1}$ and $\beta_{R1}$ are the log hazard ratios of treatment on the nonterminal event time and gap time, respectively, for subjects with $U=1$; the parameters $\beta_{M2}$ and $\beta_{R2}$ are the log hazard ratios on the nonterminal event time and gap time, respectively, comparing subjects with $U=1$ and $U=2$ who both unexposed to treatment with baseline covariates value $\bX =\bzero$.

For subject with $U=2$ and exposed with treatment ($A=1$), we assume that the terminal event time follows a proportional hazards model with hazard function given by
\[\lambda_T^{(2)}(t|A=1, \bX=\bx) =\lambda_3(t)\exp\left(\beta_{T2} + \bgamma_{T2}^{\rm T}\bx\right).\]
For subject with $U=3$, we suppose that the terminal event time follows a proportional hazards model with hazard function given by
\[\lambda_T^{(3)}(t|A=a, \bX=\bx) =\lambda_3(t)\exp\left(\beta_{T3} a + \bgamma_{T3}^{\rm T}\bx\right).\]
Note that the terminal event times for subject with $U=3$ and subject with $U=2$ exposed to treatment share the same baseline hazard function.
The parameter $\beta_{T3}$ is the log hazard ratio of treatment on the terminal event time for subjects with $U=3$, while $\beta_{T2}$ is the log hazard ratio of the terminal event time comparing subjects with $U=3$ and $A=0$ with subjects with $U=2$ and $A=1$, with the same covariates value $\bX=\bzero$.

%\begin{remark}
%Current semiparametric model further constraints the stratum-specific distributions in strata with $U=1$ and $U=3$, comparing to Assumption \ref{ass:model}.
%That is, we further assume that given $U=1$ the hazard functions corresponding to $\{M(0),T(0)\}$ and $\{M(1),T(1)\}$ are proportional, and given $U=3$ the hazard functions corresponding to $T(0)$ and $T(1)$ are proportional.
%The proportionality assumption conforms to common practice, and it increases power and stability in parameter estimation.
%\end{remark}

The natural indirect and direct effects in stratum with $U=1$ can be presented as
\begin{align*}
NIE_1(t|\bX = \bx) = &\int_0^t \exp\left\{-\Lambda_2(t-m)e^{\beta_{R1}+\bgamma_{R1}^{\rm T}\bx}\right\}\lambda_1(m)e^{\bgamma_{M1}^{\rm T}\bx}\\
&\qquad\times\left[e^{\beta_{M1}}\exp\left\{-\Lambda_1(m)e^{\beta_{M1} + \bgamma_{M1}^{\rm T}\bx}\right\} - \exp\left\{-\Lambda_1(m)e^{\bgamma_{M1}^{\rm T}\bx}\right\}\right]dm\\
&\qquad + \exp\left\{-\Lambda_1(t)e^{\beta_{M1} + \bgamma_{M1}^{\rm T}\bx}\right\} - \exp\left\{-\Lambda_1(t)e^{\bgamma_{M1}^{\rm T}\bx}\right\}
\end{align*}
and
\begin{align*}
NDE_1(t|\bX = \bx) =&\int_0^t 
\left[\exp\left\{-\Lambda_2(t-m)e^{\beta_{R1} + \bgamma_{R1}^{\rm T}\bx}\right\} - \exp\left\{-\Lambda_2(t-m)e^{\bgamma_{R1}^{\rm T}\bx}\right\}\right]\\
&\qquad\times \lambda_1(m)e^{\bgamma_{M1}^{\rm T}\bx} \exp\left\{-\Lambda_1(m)e^{\bgamma_{M1}^{\rm T}\bx}\right\}d m,
\end{align*}
where $\Lambda_1(t) = \int_0^t\lambda_1(s)ds$ and $\Lambda_2(t) = \int_0^t\lambda_2(s)ds$.
The total effects in strata with $U=2$ and $U=3$ are given by
\begin{align*}
TE_2(t|\bX = \bx) = &\exp\left\{-\Lambda_3(t)e^{\beta_{T2} + \bgamma_{T2}^{\rm T}\bx}\right\} - 1 + \int_0^t \lambda_1(m)e^{\beta_{M2} + \bgamma_{M2}^{\rm T}\bx} \\
&\qquad\times\exp\left\{-\Lambda_1(m)e^{\beta_{M2} + \bgamma_{M2}^{\rm T}\bx}\right\}\left[1 - \exp\left\{-\Lambda_2(t-m)e^{\beta_{R2} + \bgamma_{R2}^{\rm T}\bx}\right\}\right]dm\end{align*}
and
\[TE_3(t|\bX = \bx) = \exp\left\{-\Lambda_3(t)e^{\beta_{T3} + \bgamma_{T3}^{\rm T}\bx}\right\} - \exp\left\{-\Lambda_3(t)e^{\bgamma_{T3}^{\rm T}\bx}\right\},\]
where $\Lambda_3(t) = \int_0^t\lambda_3(s)ds$.

As in \cite{yu2015semiparametric}, we consider a multinomial logistic regression model on the stratum membership.
In particular, we assume
\begin{align*}
w_1(\bx; \balpha) = \Pr(U=1|\bX=\bx) = \frac{\exp\left(\balpha_1^{\rm T}\tbx\right)}{1 + \exp\left(\balpha_1^{\rm T}\tbx\right) + \exp\left(\balpha_2^{\rm T}\tbx\right)},\\
w_2(\bx; \balpha) = \Pr(U=2|\bX=\bx) = \frac{\exp\left(\balpha_2^{\rm T}\tbx\right)}{1 + \exp\left(\balpha_1^{\rm T}\tbx\right) + \exp\left(\balpha_2^{\rm T}\tbx\right)},
\end{align*}
and $w_3(\bx; \balpha) = \Pr(U=3|\bX=\bx) = \{1 + \exp\left(\balpha_1^{\rm T}\tbx\right) + \exp\left(\balpha_2^{\rm T}\tbx\right)\}^{-1}$, where $\balpha = (\balpha_1^{\rm T},\balpha_2^{\rm T})^{\rm T}$ and $\tbx = (1,\bx^{\rm T})^{\rm T}$.
Then, the marginalized stratum-specific natural indirect and direct effects are given by
\begin{align*}NIE_1(t) &=Pr\{T(1,M(1))\geq t|U=1\}-Pr\{T(1,M(0))\geq t|U=1\}\\&= \frac{\int NIE_1(t|\bX = \bx)w_1(\bx; \balpha)dF(\bx)}{\int w_1(\bx; \balpha)dF(\bx)}
\end{align*}
and
\begin{align*}
NDE_1(t) 
&= Pr\{T(1,M(0))\geq t|U=1\}-Pr\{T(0,M(0))\geq t|U=1\}
\\&= \frac{\int NDE_1(t|\bX = \bx)w_1(\bx; \balpha)dF(\bx)}{\int w_1(\bx; \balpha)dF(\bx)},
\end{align*}
where $F(\cdot)$ is the cumulative distribution function of $\bX$.

\subsection{Nonparametric Maximum Likelihood Estimation}
For a random sample of $n$ subjects, the observed semi-competing risks data are given by $\calO = \{\calO_i:i=1,\dots,n\}$, where 
\[\calO_i = \{\Delta_i^M, Z_i, \Delta_i^T, Y_i, A_i, \bX_i\}.\]
%$\Delta_i^M$ and $\Delta_i^T$ are the indicators of occurrences of the nonterminal and terminal events, respectively, $Y_i\equiv \min(T_i, C_i)$ and $Z_i = \min(M_i,Y_i)$ are the observation times for the two events, and $C_i$ is a censoring time conditional independent of $M_i$ and $T_i$ given $A_i$ and $\bX_i$.
For $i=1,\dots,n$, if $\Delta_i^M = 1$, then the likelihood corresponding to subject $i$ is given by
\begin{align*}
\tL_{i1}(\calO_i) =&\Pr\left(U_i=1|\bX_i\right)\Pr\left(Z_i,Y_i,\Delta_i^T|U_i=1,\bX_i,A_i\right)\\
&\qquad + I\left(A_i=0\right)\Pr\left(U_i=2|\bX_i\right)\Pr\left(Z_i,Y_i,\Delta_i^T|U_i=2,\bX_i,A_i=0\right);
\end{align*}
if $\Delta_i^M = 0$ and $\Delta_i^T=1$, then the likelihood corresponding to subject $i$ is given by
\begin{align*}
\tL_{i2}(\calO_i) =&\Pr\left(U_i=3|\bX_i\right)\Pr\left(Y_i,\Delta_i^T|U_i=3,\bX_i\right)\\
&\qquad + I\left(A_i=1\right)\Pr\left(U_i=2|\bX_i\right)\Pr\left(Y_i,\Delta_i^T|U_i=2,\bX_i,A_i=1\right);
\end{align*}
and if $\Delta_i^M = \Delta_i^T = 0$, then the likelihood corresponding to subject $i$ is given by
\begin{align*}
\tL_{i3}(\calO_i) =&\Pr\left(U_i=1|\bX_i\right)\Pr\left(Z_i,\Delta_i^M, Y_i,\Delta_i^T|U_i=1,\bX_i\right)\\
&\qquad + \Pr\left(U_i=2|\bX_i\right) \left\{I\left(A_i=0\right)\Pr\left(Z_i,\Delta_i^M, Y_i,\Delta_i^T|U_i=2,\bX_i,A_i=0\right) \right.\\
&\qquad\qquad\left. + I\left(A_i=1\right)\Pr\left(Y_i,\Delta_i^T|U_i=2,\bX_i,A_i=1\right)\right\}\\
&\qquad + \Pr\left(U_i=3|\bX_i\right)\Pr\left(Y_i,\Delta_i^T|U_i=3,\bX_i\right).
\end{align*}
Therefore, the likelihood function for the observed data $\calO$ is given by
\[\prod_{i=1}^n\tL_{i1}(\calO_i)^{\Delta_i^M}\left\{\tL_{i2}(\calO_i)^{\Delta_i^T}\tL_{i3}(\calO_i)^{1-\Delta_i^T}\right\}^{1-\Delta_i^M}.\]

We consider the nonparametric maximum likelihood estimation such that the estimators for $\Lambda_1$, $\Lambda_2$, and $\Lambda_3$ are step functions.
In particular, let $0<t_{11}<\dots<t_{1m_1}<\infty$ be the ordered sequence of event times $Z_i$'s with $\Delta_i^M = 1$; let $0<t_{21}<\dots<t_{2m_2}<\infty$ be the ordered sequence of gap times $V_i \equiv Y_i-Z_i$'s with $\Delta_i^M = \Delta_i^T = 1$; and let $0<t_{31}<\dots<t_{3m_3}<\infty$ be the ordered sequence of event times $Y_i$'s with $\Delta_i^M = 0$ and $\Delta_i^T = 1$.
Let $\lambda_{kl}$ be the jump size for $\Lambda_k$ at $t_{kl}$ for $k=1, 2, 3$ and $l=1,\dots, m_k$.
Write $\bfeta_{M1} = (\beta_{M1},\bgamma_{M1})^{\rm T}$, $\bfeta_{R1} = (\beta_{R1},\bgamma_{R1})^{\rm T}$, $\bfeta_{M2} = (\beta_{M2},\bgamma_{M2})^{\rm T}$, $\bfeta_{R2} = (\beta_{R2},\bgamma_{R2})^{\rm T}$, $\bfeta_{T2} = (\beta_{T2},\bgamma_{T2})^{\rm T}$, $\bfeta_{T3} = (\beta_{T3},\bgamma_{T3})^{\rm T}$, $\btheta = (\bfeta_{M1}^{\rm T},\bfeta_{R1}^{\rm T},\bfeta_{M2}^{\rm T},\bfeta_{R2}^{\rm T},\bfeta_{T2}^{\rm T},\bfeta_{T3}^{\rm T},\balpha^{\rm T})^{\rm T}$, and $\calA = (\Lambda_1,\Lambda_2,\Lambda_3)^{\rm T}$.
We maximize the objective function 
\[L_n(\btheta,\calA) = \prod_{i=1}^nL_{i1}(\btheta,\calA)^{\Delta_i^M}\left\{L_{i2}(\btheta,\calA)^{\Delta_i^T}L_{i3}(\btheta,\calA)^{1-\Delta_i^T}\right\}^{1-\Delta_i^M},\]
where 
\begin{align*}
L_{i1}(\btheta,\calA)=&w_1(\bX_i; \balpha)\Lambda_1\{Z_i\}e^{\bfeta_{M1}^{\rm T}\bW_i}\exp\left(-e^{\bfeta_{M1}^{\rm T}\bW_i}\sum_{t_{1l}\le Z_i}\lambda_{1l}\right)\\
&\qquad\times\left(\Lambda_2\{V_i\}e^{\bfeta_{R1}^{\rm T}\bW_i}\right)^{\Delta_i^T}\exp\left(-e^{\bfeta_{R1}^{\rm T}\bW_i}\sum_{t_{2l}\le V_i}\lambda_{2l}\right)\\
&+ I(A_i=0)w_2(\bX_i; \balpha)\Lambda_1\{Z_i\}e^{\bfeta_{M2}^{\rm T}\tbX_i}\exp\left(-e^{\bfeta_{M2}^{\rm T}\tbX_i}\sum_{t_{1l}\le Z_i}\lambda_{1l}\right)\\
&\qquad\times\left(\Lambda_2\{V_i\}e^{\bfeta_{R2}^{\rm T}\tbX_i}\right)^{\Delta_i^T}\exp\left(-e^{\bfeta_{R2}^{\rm T}\tbX_i}\sum_{t_{2l}\le V_i}\lambda_{2l}\right),\\
L_{i2}(\btheta,\calA)=&I(A_i=1)w_2(\bX_i; \balpha)\left(\Lambda_3\{Y_i\}e^{\bfeta_{T2}^{\rm T}\tbX_i}\right)^{\Delta_i^T}\exp\left(-\sum_{t_{3l}\le Y_i}\lambda_{3l}e^{\bfeta_{T2}^{\rm T}\tbX_i}\right)\\
&\qquad + w_3(\bX_i; \balpha)\left(\Lambda_3\{Y_i\}e^{\bfeta_{T3}^{\rm T}\bW_i}\right)^{\Delta_i^T}\exp\left(-\sum_{t_{3l}\le Y_i}\lambda_{3l}e^{\bfeta_{T3}^{\rm T}\bW_i}\right),\\
L_{i3}(\btheta,\calA)=&L_{i2}(\bfeta,\calA) +w_1(\bX_i; \balpha)\exp\left(-\sum_{t_{1l}\le Z_i}\lambda_{1l}e^{\bfeta_{M1}^{\rm T}\bW_i}\right)\\
&\qquad+ I(A_i=0)w_2(\bX_i; \balpha)\exp\left(-\sum_{t_{1l}\le Z_i}\lambda_{1l}e^{\bfeta_{M2}^{\rm T}\tbX_i}\right),
\end{align*}
$\bW_i = (A_i,\bX_i^{\rm T})^{\rm T}$, and $\Lambda_k\{t\}$ is the jump size of $\Lambda_k$ at time $t$ for $k=1,2,3$.

By treating $U_i$ ($i=1, \dots, n$) as missing data, we propose an EM algorithm to maximize this objective function.
The details of the EM algorithm are given in Appendix B.
We write $(\wbtheta, \wcalA)$ as the estimators.
The indirect and direct effects in stratum with $U=1$ can then be estimated by
\begin{align}
&\widehat{NIE}_1(t;\bx) = \sum_{t_{1j}\le t}\left[
\exp\left(-\sum_{t_{2k}\le t-t_{1j}}\wlambda_{2k}e^{\wbtheta_{R1}^{\rm T}\tbx }\right)\wlambda_{1j}\right.\nonumber\\
&\qquad\left.\times\left\{e^{\wbtheta_{M1}^{\rm T}\tbx}\exp\left(-\sum_{k=1}^j\wlambda_{1k}e^{\wbtheta_{M1}^{\rm T}\tbx}\right) - e^{\wbgamma_{M1}^{\rm T}\bx}\exp\left(-\sum_{k=1}^j\wlambda_{1k}e^{\wbgamma_{M1}^{\rm T}\bx}\right)\right\}\right]\nonumber\\
&\qquad + \exp\left(-\sum_{t_{1j}\le t}\wlambda_{1j}e^{\wbtheta_{M1}^{\rm T}\tbx}\right) - \exp\left(-\sum_{t_{1j}\le t}\wlambda_{1j}e^{\wbgamma_{M1}^{\rm T}\bx}\right)\label{equ:NIE1_est}
\end{align}
and
\begin{align}
\widehat{NDE}_1(t;\bx) =& \sum_{t_{1j}\le t}\left[
\left\{\exp\left(-\sum_{t_{2k}\le t-t_{1j}}\wlambda_{2k}e^{\wbtheta_{R1}^{\rm T}\tbx}\right) - \exp\left(-\sum_{t_{2k}\le t-t_{1j}}\wlambda_{2k}e^{\wbgamma_{R1}^{\rm T}\bx}\right)\right\}\right.\nonumber\\
&\qquad\left.\times \wlambda_{1j}e^{\wbgamma_{M1}^{\rm T}\bx} \exp\left(-\sum_{k=1}^j\wlambda_{1k}e^{\wbgamma_{M1}^{\rm T}\bx}\right)\right].\label{equ:NDE1_est}
\end{align}
The total effects in strata with $U=2$ and $U=3$ can be estimated by
\begin{align}
&\widehat{TE}_2(t;\bx) = \exp\left(-\sum_{t_{3j}\le t}\wlambda_{3j}e^{\wbtheta_{T2}^{\rm T}\tbx}\right) - 1\nonumber\\
& \qquad + \sum_{t_{1j}\le t}\left[\wlambda_{1j}e^{\wbtheta_{M2}^{\rm T}\tbx}\exp\left(-\sum_{k=1}^j\wlambda_{1k}e^{\wbtheta_{M2}^{\rm T}\tbx} \right) \left\{1 - \exp\left(-\sum_{t_{2k}\le t-t_{1j}}\wlambda_{2k}e^{\wbtheta_{R2}^{\rm T}\tbx}\right)\right\}\right]\label{equ:TE2_est}
\end{align}
and
\begin{equation}
 \widehat{TE}_3(t;\bx) = \exp\left(-\sum_{t_{3j}\le t}\wlambda_{3j}e^{\wbtheta_{T3}^{\rm T}\tbx}\right) - \exp\left(-\sum_{t_{3j}\le t}\wlambda_{3j}e^{\wbgamma_{T3}^{\rm T}\bx}\right).\label{equ:TE3_est}
\end{equation}
The marginalized stratum-specific indirect and direct effects in stratum with $U=1$ can be estimated by
\begin{equation}
\widehat{NIE}_1(t) = \frac{\sum_{i=1}^nw_1(\bX_i;\wbalpha)\widehat{NIE}_1(t;\bX_i)}{\sum_{i=1}^nw_1(\bX_i;\wbalpha)}\end{equation}
and
\begin{equation}\widehat{NDE}_1(t) = \frac{\sum_{i=1}^nw_1(\bX_i;\wbalpha)\widehat{NDE}_1(t;\bX_i)}{\sum_{i=1}^nw_1(\bX_i;\wbalpha)},\label{equ:NDE_last}
\end{equation}
respectively.

\subsection{Asymptotic Properties}
We study the asymptotic properties of the estimators under the semiparametric model in Section \ref{sec:PHmodel}.
Under suitable regularity conditions, the estimators $(\wbtheta,\wcalA)$ has the usual large sample properties, including consistency and asymptotic normality, as given in Theorem \ref{thm:para} below.
Let $\btheta_0$, $\Lambda_{10}$, $\Lambda_{20}$, and $\Lambda_{30}$ be the true values of $\btheta$, $\Lambda_1$, $\Lambda_2$, and $\Lambda_3$, respectively, $\|\cdot\|$ be the Euclidean norm, and $\tau_k$ be the upper limit of the support of $\wLambda_k$ for $k=1,2,3$.

\begin{thm}\label{thm:para}
Under Conditions 1-5 in Appendix C,
\[\left\|\wbtheta - \btheta_0\right\| + \sum_{k=1}^3\sup_{t\in[0,\tau_k]}\left|\wLambda_k(t) - \Lambda_{k0}(t)\right| \]
converges to zero almost surely.
In addition, $\sqrt n \{\wbtheta - \btheta_0, \wLambda_1(\cdot) - \Lambda_{10})(\cdot), \wLambda_2(\cdot) - \Lambda_{20})(\cdot), \wLambda_3(\cdot) - \Lambda_{30})(\cdot)\}$ converges weakly to a zero-mean Gaussian process in the Banach space $\mathR^m\times l^\infty(\calA_1)\times l^\infty(\calA_2)\times l^\infty(\calA_3)$, where $m$ is the dimension of $\btheta$ and $\calA_k$ is the unit ball in the space of functions on $[0,\tau_k]$ with bounded variation for $k=1,2,3$.
\end{thm}
\begin{thm}\label{thm:med}
Under Conditions 1-5 in Appendix C, the estimators for stratum-specific effects given in (\ref{equ:NIE1_est})-(\ref{equ:NDE_last}) are consistent and asymptotically normal.
\end{thm}
The proofs of Theorems \ref{thm:para} and \ref{thm:med} are given in Appendix C.
Since the form of the limiting variances of the stratum-specific effects is complicated, we estimate the variance of the estimators by a nonparametric bootstrap procedure in all numerical studies.

\section{Simulation Studies}
We conducted simulation studies to examine the performance of the proposed methods.
We generated two covariates $X_1\sim N(0, 1)$ and $X_2 \sim Unif (0, 1)$ and generated the treatment indicator $A\sim Bin(0.5)$ to reflect 1:1 randomization.
We set $\Lambda_1(t) = t$, $\Lambda_2(t) = 0.2t$, and $\Lambda_3(t) = \log (1 + t)$, while the true values of the other parameters are shown in Table 1, along with the simulation results.
We generated a censoring time $C\sim Unif(0, 15)$ to obtain approximately 51\% and 26\% censoring rates for the nonterminal and terminal events, respectively.
The proportions of subjects with $U=1, 2, 3$ are approximately 31\%, 41\%, and 28\%, respectively.

We considered $1000$ replicates with sample sizes $n=1000$ and $2000$, where $100$ bootstrap samples were used for variance estimation.
Table 1 shows the simulation results, where Bias, SE and SEE denote, respectively, the averaged bias, empirical standard error and averaged standard error estimates, and CP stands for the empirical coverage probability of the 95\% confidence intervals.
All examined replications converge with a $10^{-6}$ convergence criterion.
The parameter estimators are virtually unbiased.
The bootstrap variance estimator overestimates the true variability for some of the parameters, but it gets more accurate when sample size increases.

\begin{table}
\protect\caption{Simulation Results for Regression Parameters}\label{tab:simu}
\medskip{}
\begin{centering}
\renewcommand*{\arraystretch}{1}
\renewcommand\tabcolsep{3pt}
 \begin{threeparttable}
{\footnotesize{}
\begin{tabular}{ccccccccccccccccc}
\hline\hline
	&	{True}	&	\multicolumn{4}{c}{$n=1000$}							&&	\multicolumn{4}{c}{$n=2000$}							\\
	&	Value	&	Bias	&	SE	&	SEE	&	CP	&&	Bias	&	SE	&	SEE	&	CP	\\\hline
$\beta_{M1}$	&$	0.5	$&$	0.008	$&$	0.224	$&$	0.261	$&$	0.97	$&&$	0.006	$&$	0.147	$&$	0.168	$&$	0.97	$\\
$\bgamma_{M1}$	&$	0.5	$&$	0.010	$&$	0.091	$&$	0.095	$&$	0.96	$&&$	0.005	$&$	0.063	$&$	0.063	$&$	0.94	$\\
	&$	0.5	$&$	0.014	$&$	0.269	$&$	0.296	$&$	0.96	$&&$	0.010	$&$	0.192	$&$	0.195	$&$	0.96	$\\
$\beta_{R1}$	&$	0.5	$&$	-0.043	$&$	0.276	$&$	0.334	$&$	0.97	$&&$	-0.025	$&$	0.161	$&$	0.202	$&$	0.97	$\\
$\bgamma_{R1}$	&$	-0.2	$&$	-0.011	$&$	0.095	$&$	0.102	$&$	0.96	$&&$	-0.004	$&$	0.065	$&$	0.067	$&$	0.95	$\\
	&$	-0.2	$&$	0.001	$&$	0.308	$&$	0.332	$&$	0.97	$&&$	0.002	$&$	0.212	$&$	0.214	$&$	0.95	$\\\\
$\beta_{M2}$	&$	-0.2	$&$	0.005	$&$	0.503	$&$	0.599	$&$	0.98	$&&$	-0.002	$&$	0.327	$&$	0.380	$&$	0.97	$\\
$\bgamma_{M2}$	&$	0.4	$&$	0.025	$&$	0.123	$&$	0.154	$&$	0.98	$&&$	0.011	$&$	0.078	$&$	0.090	$&$	0.96	$\\
	&$	0.5	$&$	0.017	$&$	0.393	$&$	0.476	$&$	0.98	$&&$	0.012	$&$	0.252	$&$	0.287	$&$	0.97	$\\
$\beta_{R2}$	&$	0.4	$&$	-0.068	$&$	0.586	$&$	0.699	$&$	0.97	$&&$	-0.028	$&$	0.358	$&$	0.435	$&$	0.96	$\\
$\bgamma_{R2}$	&$	0.5	$&$	-0.013	$&$	0.187	$&$	0.230	$&$	0.96	$&&$	-0.002	$&$	0.115	$&$	0.138	$&$	0.97	$\\
	&$	0.5	$&$	-0.020	$&$	0.476	$&$	0.546	$&$	0.97	$&&$	-0.002	$&$	0.303	$&$	0.337	$&$	0.96	$\\
$\beta_{T2}$	&$	0.0	$&$	0.028	$&$	0.485	$&$	0.523	$&$	0.98	$&&$	0.011	$&$	0.357	$&$	0.359	$&$	0.96	$\\
$\bgamma_{T2}$	&$	-0.5	$&$	0.015	$&$	0.175	$&$	0.197	$&$	0.98	$&&$	0.012	$&$	0.124	$&$	0.132	$&$	0.97	$\\
	&$	-0.2	$&$	-0.018	$&$	0.435	$&$	0.509	$&$	0.98	$&&$	0.009	$&$	0.312	$&$	0.324	$&$	0.95	$\\\\
$\beta_{T3}$	&$	0.2	$&$	-0.044	$&$	0.410	$&$	0.448	$&$	0.98	$&&$	-0.034	$&$	0.334	$&$	0.332	$&$	0.96	$\\
$\bgamma_{T3}$	&$	-0.2	$&$	-0.035	$&$	0.107	$&$	0.118	$&$	0.95	$&&$	-0.024	$&$	0.076	$&$	0.078	$&$	0.95	$\\
	&$	0.0	$&$	-0.017	$&$	0.371	$&$	0.402	$&$	0.96	$&&$	-0.029	$&$	0.265	$&$	0.261	$&$	0.95	$\\\\
$\balpha_1$	&$	0.0	$&$	0.009	$&$	0.199	$&$	0.213	$&$	0.96	$&&$	0.010	$&$	0.148	$&$	0.146	$&$	0.95	$\\
	&$	0.3	$&$	-0.005	$&$	0.113	$&$	0.113	$&$	0.95	$&&$	-0.003	$&$	0.078	$&$	0.078	$&$	0.95	$\\
	&$	0.1	$&$	-0.002	$&$	0.343	$&$	0.367	$&$	0.96	$&&$	-0.007	$&$	0.256	$&$	0.253	$&$	0.95	$\\
$\balpha_2$	&$	0.2	$&$	0.025	$&$	0.296	$&$	0.321	$&$	0.96	$&&$	0.010	$&$	0.201	$&$	0.211	$&$	0.96	$\\
	&$	-0.5	$&$	-0.019	$&$	0.155	$&$	0.175	$&$	0.97	$&&$	-0.019	$&$	0.108	$&$	0.114	$&$	0.95	$\\
	&$	0.3	$&$	-0.023	$&$	0.491	$&$	0.537	$&$	0.97	$&&$	0.000	$&$	0.342	$&$	0.355	$&$	0.96	$\\
\hline
\end{tabular}}
 \begin{tablenotes}
 \footnotesize
 \item NOTE: Bias, SE and SEE denote, respectively, the mean bias, empirical standard error and mean standard error estimator.
 CP stands for the empirical coverage probability of the 95\% confidence interval.
 \end{tablenotes}
 \end{threeparttable}
\par\end{centering}
\protect
\end{table}

\begin{figure}[h]
\centering
\includegraphics[width=6.5in]{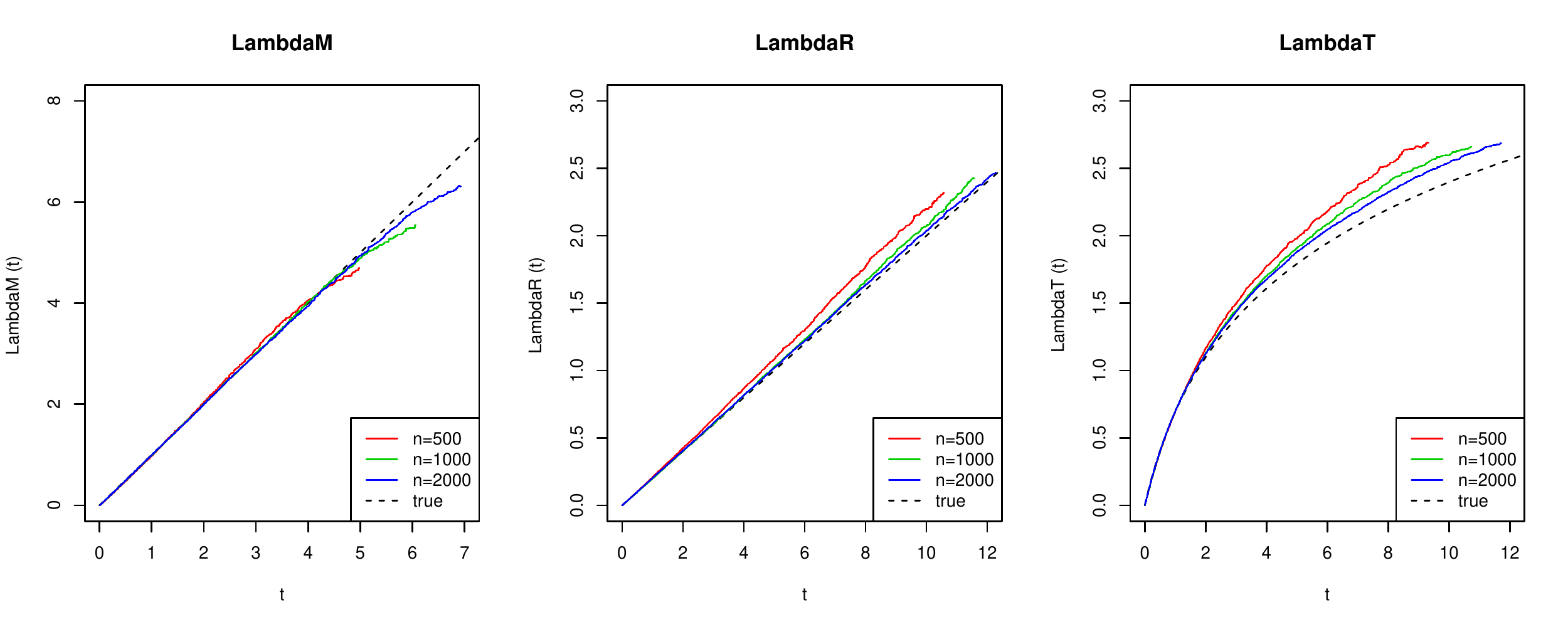}
\caption{Performance of the estimated baseline cumulative hazard functions.}\label{fig:simulambda}
\end{figure}

The estimators for the baseline cumulative hazard functions only take jump till the last observation times, such that the estimates after the last observation time is not meaningful.
Therefore, to summarize the performance of the baseline hazard estimators, for every time point $t$, we only consider the replicates with last observation time greater than $t$.
Figure \ref{fig:simulambda} shows the median of the estimated baseline hazard functions, among such replicates.
We plot till the time point at which at least 800 replicates have meaningful estimates.
The bias gets smaller as sample size increases.

Table 2 shows the performance of the estimated stratum-specific indirect and direct effects in stratum with $U=1$ and $\bX = (0.5,0.5)^{\rm T}$, as well as the estimated total effects for strata with $U=2, 3$ and  the same covariate values.
Similarly, for any $t$ the average was taken over all the replicates with estimators that have last jump time no less than $t$.
The bias gets smaller as sample size increases.
The variance estimator is accurate and the coverage probability is close to the nominal level when sample size is large.
\begin{table}
\protect\caption{Simulation Results for Stratum-Specific Mediation Effects and Total Effects}\label{tab:simu_med}
\medskip{}
\begin{centering}
\renewcommand*{\arraystretch}{1}
\renewcommand\tabcolsep{3pt}
 \begin{threeparttable}
{\footnotesize{}
\begin{tabular}{ccccccccccccccccc}
\hline\hline
	&		&	{True}	&	\multicolumn{4}{c}{$n=1000$}							&&	\multicolumn{4}{c}{$n=2000$}							\\
	&	t	&	Value	&	Bias	&	SE	&	SEE	&	CP	&&	Bias	&	SE	&	SEE	&	CP	\\\hline
$NDE_1$	&	2	&$	-0.11	$&$	0.014	$&$	0.056	$&$	0.070	$&$	0.97	$&&$	0.007	$&$	0.031	$&$	0.041	$&$	0.97	$\\
	&	4	&$	-0.17	$&$	0.018	$&$	0.090	$&$	0.106	$&$	0.96	$&&$	0.010	$&$	0.052	$&$	0.066	$&$	0.97	$\\
	&	6	&$	-0.18	$&$	0.020	$&$	0.096	$&$	0.107	$&$	0.94	$&&$	0.010	$&$	0.057	$&$	0.069	$&$	0.96	$\\\\
$NIE_1$	&	2	&$	-0.04	$&$	-0.001	$&$	0.022	$&$	0.025	$&$	0.97	$&&$	0.000	$&$	0.015	$&$	0.017	$&$	0.97	$\\
	&	4	&$	-0.03	$&$	-0.001	$&$	0.016	$&$	0.018	$&$	0.97	$&&$	-0.001	$&$	0.011	$&$	0.012	$&$	0.96	$\\
	&	6	&$	-0.02	$&$	-0.001	$&$	0.010	$&$	0.011	$&$	0.97	$&&$	0.000	$&$	0.007	$&$	0.007	$&$	0.96	$\\\\
$TE_2$	&	2	&$	-0.10	$&$	-0.036	$&$	0.158	$&$	0.175	$&$	0.97	$&&$	-0.022	$&$	0.115	$&$	0.126	$&$	0.97	$\\
	&	4	&$	0.10	$&$	-0.046	$&$	0.159	$&$	0.180	$&$	0.96	$&&$	-0.026	$&$	0.113	$&$	0.127	$&$	0.97	$\\
	&	6	&$	0.17	$&$	-0.047	$&$	0.136	$&$	0.156	$&$	0.97	$&&$	-0.027	$&$	0.097	$&$	0.109	$&$	0.97	$\\
	&	8	&$	0.18	$&$	-0.044	$&$	0.115	$&$	0.131	$&$	0.96	$&&$	-0.025	$&$	0.084	$&$	0.093	$&$	0.96	$\\\\
$TE_3$	&	2	&$	-0.07	$&$	0.021	$&$	0.139	$&$	0.147	$&$	0.97	$&&$	0.015	$&$	0.117	$&$	0.114	$&$	0.95	$\\
	&	4	&$	-0.06	$&$	0.031	$&$	0.119	$&$	0.126	$&$	0.98	$&&$	0.022	$&$	0.101	$&$	0.099	$&$	0.96	$\\
	&	6	&$	-0.06	$&$	0.033	$&$	0.101	$&$	0.107	$&$	0.97	$&&$	0.024	$&$	0.086	$&$	0.085	$&$	0.96	$\\
	&	8	&$	-0.05	$&$	0.033	$&$	0.088	$&$	0.093	$&$	0.97	$&&$	0.024	$&$	0.075	$&$	0.074	$&$	0.96	$\\
\hline
\end{tabular}}
 \begin{tablenotes}
 \footnotesize
 \item NOTE: Bias, SE and SEE denote, respectively, the mean bias, empirical standard error and mean standard error estimator.
 CP stands for the empirical coverage probability of the 95\% confidence interval.
 \end{tablenotes}
 \end{threeparttable}
\par\end{centering}
\protect
\end{table}

\section{Application}
We consider application of the proposed methods to a prostate cancer clinical trial.
NCIC Clinical Trials Group PR.3/Medical Research Council PR07/Intergroup T94-0110 is a randomized controlled trial of patients with locally advanced prostate cancer.
The primary objective is to determine whether the addition of radiotherapy (RT) to androgen-deprivation therapy (ADT) prolonged overall survival, defined as time from random assignment to death from any cause.
One thousand two hundred and five patients with locally advanced prostate cancer were recruited and randomly assigned between 1995 and 2005, 602 to ADT alone and 603 to ADT + RT.
These patients were either with T3-4, N0/Nx, M0 prostate cancer or with T1-2 disease with either prostate-specific antigen (PSA) of more than 40$\mu$g/L or PSA of 20 to 40$\mu$g/L plus Gleason score of 8 to 10.
In the final report of the study \citep{mason2015final}, at a median follow-up time of 8 years, 465 patients had died.
Overall survival was significantly improved in the patients allocated to ADT + RT (hazard ratio 0.70 with 95\% CI, 0.57 to 0.85; P$<$.001).

In addition to the primary outcome of death, the study also collected data on time to disease progression, which was defined as the first of any of the following events: biochemical progression, local progression, or development of metastatic disease.
We analyzed the data to reveal the proportions of the treatment effect on overall survival that are mediated by disease progression.
Particularly, we adjusted for initial PSA level ($<$ 20 vs. 20 to 50, vs. $>$50g/L) and Gleason score(8 vs. 8 to 10). 
%We first excluded 26 subjects with the same disease progression time and death time.
%Those subjects probably were determined disease progression due to death from prostate cancer, such that times of ``disease progression'' are not accurately determined.
%We also excluded 8 subjects with missing covariate values to obtain a sample size of 1171.

We analyzed the data using the proposed approach, with 100 bootstrap samples for variance estimation.
The parameter estimates for regression coefficients for the event time processes are shown in Table \ref{tab:data7}.
For stratum with $U=1$, ADT + RT is associated with a decreased risk of disease progression, while it is associated with an increased risk from disease progression to death.
For stratum with $U=3$, ADT + RT is associated with a decreased risk of death.
The effects are not significant at 0.05 level.
For stratum with $U=1$, a subject with initial PSA level $>$50 g/L is associated with significantly increased risk of disease progression, compared to a similar subject with initial PSA level $<$20 g/L; and a subject with Gleason score 8-10 is associated with significantly decreased risk of disease progression, compared to a similar subject with Gleason score $<$8.

\begin{table}
\protect\caption{Parameter Estimates for Regression Coefficients for Event Time Processes}\label{tab:data7}
\medskip{}
\begin{centering}
\renewcommand*{\arraystretch}{1}
\renewcommand\tabcolsep{3pt}
 \begin{threeparttable}
{\footnotesize{}
\begin{tabular}{l|ccc|ccc}
\hline\hline
\multirow{2}{*}{Process}	&	\multicolumn{6}{c}{$U=1$}											\\\cline{2-7}
	&	\multicolumn{3}{c|}{Health $\to$ Disease}					&	\multicolumn{3}{c}{Disease $\to$ Death}					\\\hline
	&	Est	&	SEE	&	$p$-value	&	Est	&	SEE	&	$p$-value	\\\hline
ADT + RT	&$	-0.825	$&$	0.987	$&$	0.403	$&$	0.460	$&$	0.658	$&$	0.484	$\\
Initial PSA Level (20 to 50 g/L)	&$	0.321	$&$	0.530	$&$	0.545	$&$	-0.097	$&$	0.305	$&$	0.751	$\\
Initial PSA Level ($>$ 50 g/L)	&$	1.607	$&$	0.566	$&$	0.005	$&$	0.065	$&$	0.342	$&$	0.848	$\\
Gleason Score (8-10)	&$	-2.008	$&$	0.413	$&$	0.0000	$&$	-0.378	$&$	0.241	$&$	0.117	$\\\hline
\multirow{2}{*}{Process}	&	\multicolumn{6}{c}{$U=2$, ADT}											\\\cline{2-7}
	&	\multicolumn{3}{c|}{Health $\to$ Disease}					&	\multicolumn{3}{c}{Disease $\to$ Death}					\\\hline
	&	Est	&	SEE	&	$p$-value	&	Est	&	SEE	&	$p$-value	\\\hline
Intercept	&$	-1.917	$&$	1.587	$&$	0.227	$&$	-0.557	$&$	3.548	$&$	0.875	$\\
Initial PSA Level (20 to 50 g/L)	&$	0.663	$&$	1.005	$&$	0.510	$&$	-0.304	$&$	0.863	$&$	0.725	$\\
Initial PSA Level ($>$ 50 g/L)	&$	1.619	$&$	0.904	$&$	0.073	$&$	-0.104	$&$	0.940	$&$	0.912	$\\
Gleason Score (8-10)	&$	0.674	$&$	0.952	$&$	0.479	$&$	-0.106	$&$	3.381	$&$	0.975	$\\\hline
\multirow{2}{*}{Process}	&	\multicolumn{3}{c|}{$U=2$, ADT + DT}					&	\multicolumn{3}{c}{$U=3$}						\\\cline{2-7}
	&	\multicolumn{3}{c|}{Health $\to$ Death}					&	\multicolumn{3}{c}{Health $\to$ Death}						\\\hline
	&	Est	&	SEE	&	$p$-value	&	Est	&	SEE	&	$p$-value		\\\hline
Intercept	&$	-3.212	$&$	7.339	$&$	0.662	$&$	-0.446	$&$	0.875	$&$	0.610		$\\
Initial PSA Level (20 to 50 g/L)	&$	1.146	$&$	5.416	$&$	0.832	$&$	-0.022	$&$	0.649	$&$	0.972		$\\
Initial PSA Level ($>$ 50 g/L)	&$	1.601	$&$	5.545	$&$	0.773	$&$	-0.883	$&$	0.641	$&$	0.168		$\\
Gleason Score (8-10)	&$	1.624	$&$	4.630	$&$	0.726	$&$	-0.514	$&$	0.486	$&$	0.290		$\\
\hline
\end{tabular}}
 \end{threeparttable}
\par\end{centering}
\protect
\end{table}

\begin{figure}
\centering \makebox{\includegraphics[width=10cm]{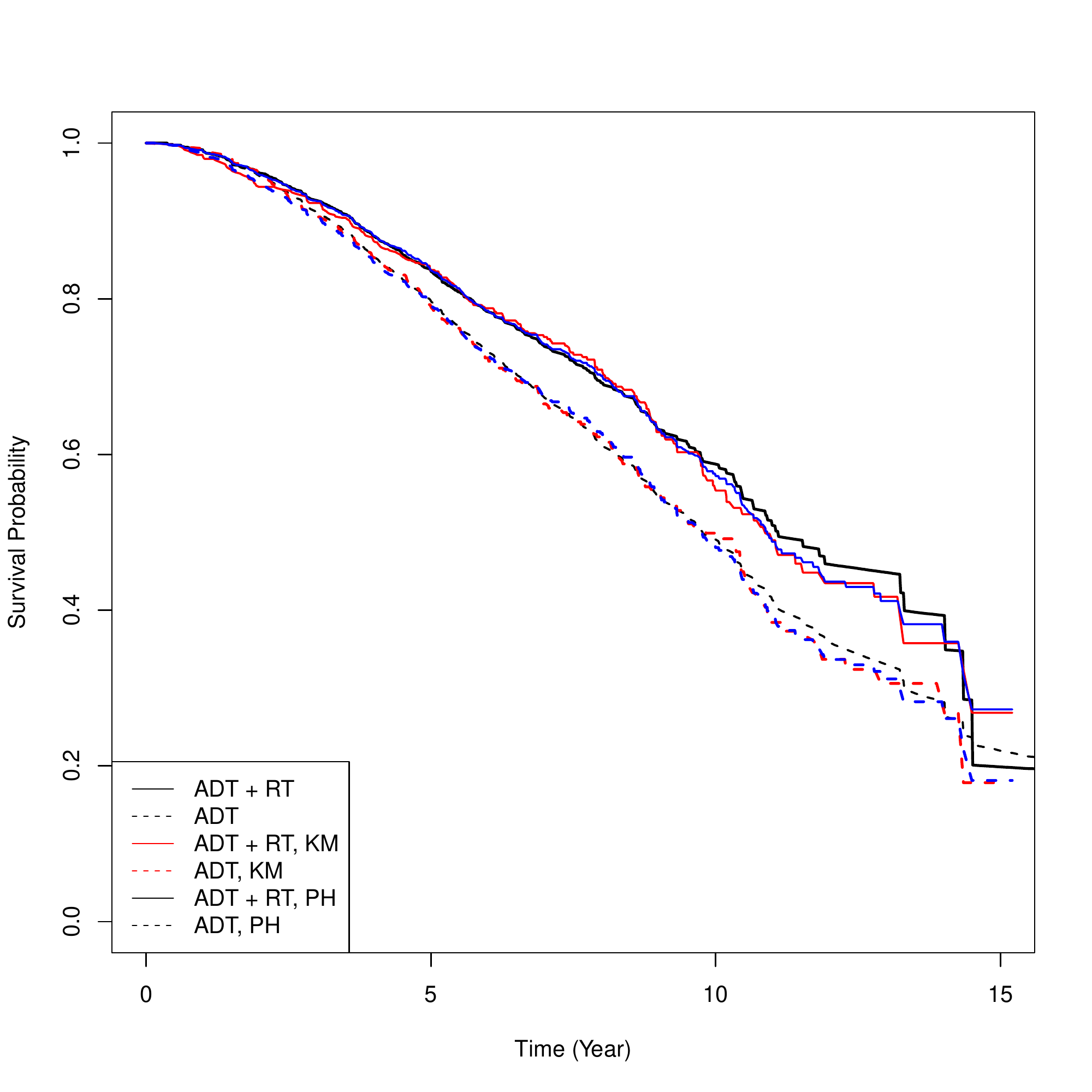}}\caption{Estimated survival functions from the proposed, Kaplan-Meier, and proportional hazards model approaches.}\label{fig:data_pop}
\end{figure}

Table \ref{tab:data7_alpha} shows the parameter estimators of the logistic regression model for stratum membership.
By averaging over the stratum membership probabilities over all subjects given their covariate values, the average probabilities of belong to strata $U=1,2$, and 3 are 40.1\%, 25.7\%, and 34.2\%, respectively.
To verify if the model is reasonable, we estimated the stratum-specific survival functions for every subject and summarize the subject-specific survival function by weighting them by his/her stratum membership probabilities.
We average the estimated survival functions for subjects assigned to ADT+RT versus ADT, and plot them against the survival function estimators from the Kaplan Meier methods and the proportional hazards model.
The results are shown in Figure \ref{fig:data_pop}.
The estimated population-average survival functions for ADT+RT and ADT groups are similar to those from the Kaplan Meier methods and the proportional hazards model, especially for time before 10 years when data are not sparse, indicating proper fit of the proposed approach.
\begin{table}
\protect\caption{Parameter Estimates for Regression Coefficients for Stratum Membership}\label{tab:data7_alpha}
\medskip{}
\begin{centering}
\renewcommand*{\arraystretch}{1}
\renewcommand\tabcolsep{3pt}
 \begin{threeparttable}
{\small{}
\begin{tabular}{l|ccc|ccc}
\hline\hline
	&	\multicolumn{3}{c|}{$\balpha_1$}					&	\multicolumn{3}{c}{$\balpha_2$}					\\\hline
	&	Est	&	SEE	&	$p$-value	&	Est	&	SEE	&	$p$-value	\\\hline
Intercept	&$	0.205	$&$	0.441	$&$	0.643	$&$	0.291	$&$	0.743	$&$	0.695	$\\
Initial PSA Level (20 to 50 g/L)	&$	0.090	$&$	0.583	$&$	0.877	$&$	0.625	$&$	1.054	$&$	0.553	$\\
Initial PSA Level ($>$ 50 g/L)	&$	-0.684	$&$	0.517	$&$	0.186	$&$	0.004	$&$	0.929	$&$	0.996	$\\
Gleason Score (8-10)	&$	0.134	$&$	0.415	$&$	0.746	$&$	-1.495	$&$	0.732	$&$	0.041	$\\
\hline
\end{tabular}}
 \end{threeparttable}
\par\end{centering}
\protect
\end{table}

\begin{figure}
\centering \makebox{\includegraphics[width=10cm]{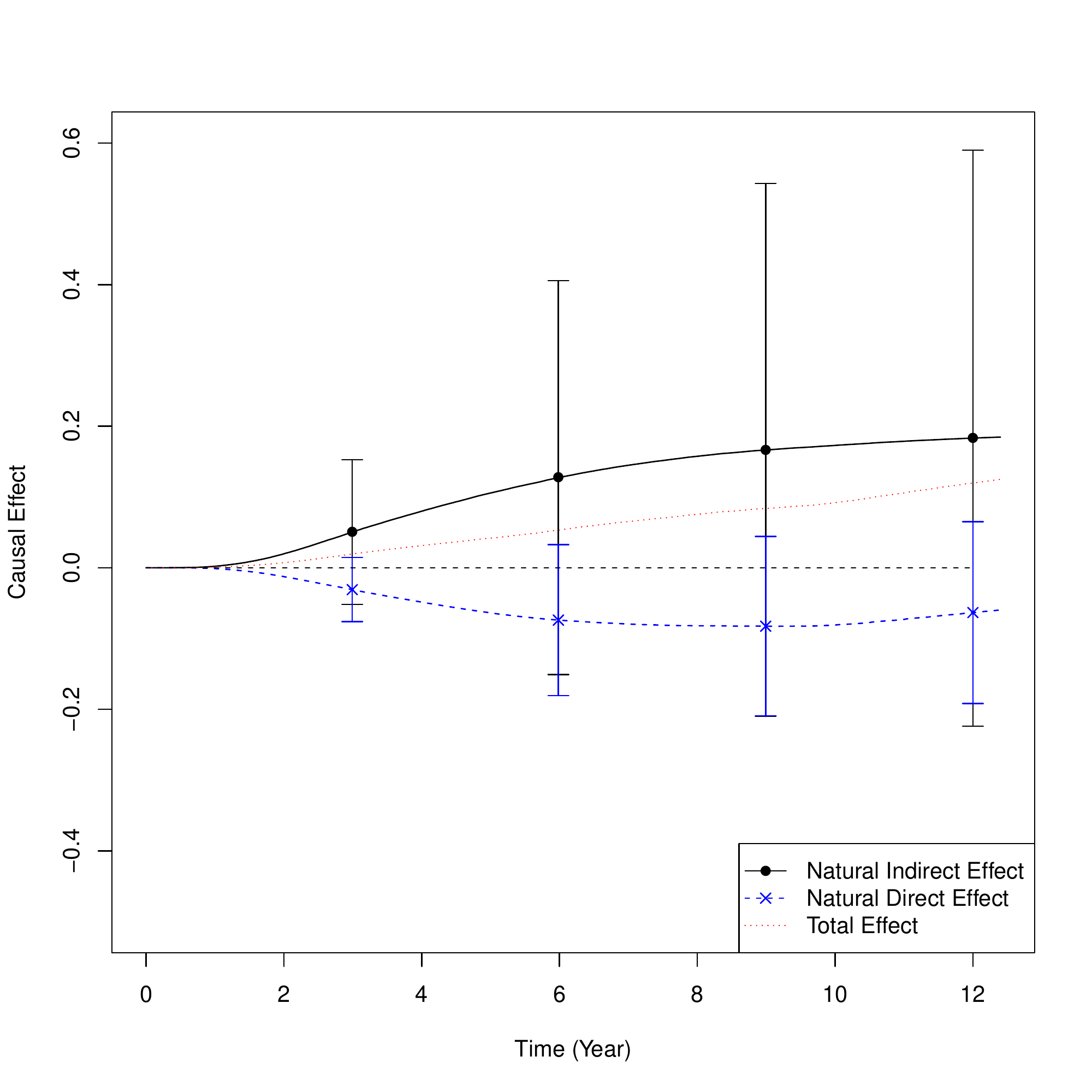}}
\caption{Estimated stratum-specific indirect and direct effects in stratum with $U=1$.}\label{fig:data_med_pop}
\end{figure}

Figure \ref{fig:data_med_pop} shows the estimated marginalized stratum-specific indirect and direct effects (with 95\% confidence intervals) for stratum with $U=1$.
The estimated natural indirect effect is positive and increasing over time, and the estimated natural direct effect is slightly negative over time.
However, the 95\% confidence intervals are wide such that the stratum-specific natural indirect and direct effects are not significant different from zero.
The total effect in stratum with $U=1$ is positive and increasing over time, corresponding to an increased survival probability assigned to ADT+RT versus ADT in stratum with $U=1$.

It is worth noting that the primary analysis for the data shows that overall survival was significantly improved in the patients allocated to ADT + RT compared to ADT.
However, our analysis failed to obtain a significant stratum-specific overall effect of ADT + RT.
The main reason is that by identifying subjects to different strata, the sample size to estimate parameters in each stratum is much smaller than that for the proportional hazards model based on all available subjects.
In addition, the proposed model has much more parameters, such that the variability for parameter estimation significantly increases.

\section{Discussion}

Semi-competing risks data are frequently observed in medical studies, where the terminal event time may censor the intermediate event time but not vice versa.
To define and estimate causal contrasts of the effect of a treatment to the terminal and intermediate events, we introduced a novel principal stratification framework that distinguishes susceptible and non-susceptible subjects given different treatments, and defined the natural indirect and direct effects in the stratum where the times to intermediate and terminal events are well-defined given both treatments.
We provided reasonable assumptions to identify the stratum-specific natural indirect and direct effects, proposed a semiparametric model, and studied an EM algorithm to obtain the nonparametric maximum likelihood estimators of model parameters.
We showed that the estimators are consistent and asymptotically efficient estimated under mild regularity conditions, and their performance are satisfactory in finite sample numerical studies.

%\fei{
%Mediation analysis with semi-competing risks data has received much attentions recently.
%In particular, \cite{huang2020causal} considered the problem and identified the natural direct and indirect effects based on assumptions on the counterfactual counting processes of the events.
%The interpretation of the effects is very different from that of our methods, in that the intermediate event is treated as time-dependent process rather than event time.}

In identifying the stratum-specific natural indirect and direct effects, we assumed that there are no subjects who are susceptible to the intermediate event under treatment ($A=1$) and non-susceptible under control ($A=0$).
This assumption may need careful examination based on scientific understanding of how treatment may affect the intermediate event.
In our data application, we assessed this assumption by fitting the proposed model with switched treatment indicator labels of ADT+RT and ADT.
The estimated probability of belonging to stratum with $U=2$ (equivalent to the fourth stratum in the original labeling) is very low, suggesting that the assumption on non-existence of the fourth stratum may be valid.
In some applications, this fourth stratum may indeed exist.
In the literature of principal stratification for uncensored data with four or more strata, the effect of interest often can only be interval identified.
Interval identification with a regression model often results in a complicated solution manifold, with properties often not well understood.
We plan to explore this problem in a future study.
%, such that treatment 1 may convert a non-susceptible subject to susceptible.
%To relax this assumption and accommodate four strata, one may impose further assumption on the model structure of the event processes, in order to (semi)-parametrically identify the stratum-specific effects.

\section*{Acknowledgement}
This manuscript was prepared using data from Dataset NCT00002633-D1 from the NCTN Data Archive of the National Cancer Institute (NCI) National Clinical Trials Network (NCTN). Data were originally collected from clinical trial NCT00002633 Phase III Randomized Trial Comparing Total Androgen Blockade Versus Total Androgen Blockade Plus Pelvic Irradiation in Clinical Stage T3-4, N0, M0 Adenocarcinoma of the Prostate All analyses and conclusions in this manuscript are the sole responsibility of the authors and do not necessarily reflect the opinions or views of the clinical trial investigators, the NCTN, or the NCI.  The authors are partially funded by the U.S. National Institutes of Health grants R01HL122212, U01AG016976 and U.S. National Science Foundation grant DMS 1711952.
Scientific computing at the Fred Hutch is supported by ORIP grant S10OD028685. 

\renewcommand{\theequation}{A.\arabic{equation}}
\setcounter{equation}{0}
\renewcommand{\thesubsection}{A.\arabic{subsection}}
\appendix
\section{Identification of Stratum-Specific Effects}
\subsection{Proof of Theorem \ref{thm:express}}\label{append:express}
Note that (\ref{equ:seq1}) in Assumption \ref{ass:seq_ign} implies
\begin{equation}
T(a, M(a^*)) \perp A | M(a^*), \bX, U=1.\label{equ:ind3}
\end{equation}
In stratum $U=1$, for a given $t\le \tau$ and any $a$, $a^*$, we have 
\begin{align*}
&\Pr\{T(a, M(a^*))\ge t |\bX=\bx, U=1 \}\\
= & \int \Pr\{T(a, M(a^*))\ge t |M(a^*) = m, \bX=\bx, U=1\}d F_{M(a^*)|\bX=\bx, U=1}(m)\\
= & \int \Pr\{T(a, M(a^*))\ge t |M(a^*) = m, A = a, \bX=\bx, U=1\}d F_{M(a^*)|\bX=\bx, U=1}(m)\\
= & \int \Pr\{T(a, M(a))\ge t |M(a) = m, A = a, \bX=\bx, U=1\}d F_{M(a^*)|\bX=\bx, U=1}(m)\\
= & \int_0^ t \Pr\{T(a)\ge t |M(a) = m, A = a, \bX=\bx, U=1\}d F_{M(a^*)|\bX=\bx, U=1}(m) \\
&\qquad + \int_ t^\infty \Pr\{T(a)\ge t |M(a) = m, A = a, \bX=\bx, U=1\}d F_{M(a^*)|\bX=\bx, U=1}(m)\\
= & \int_0^t \Pr\{T(a)\ge t |M(a) = m, A = a, \bX=\bx, U=1\}d F_{M(a^*)|\bX=\bx, U=1}(m)\\
&\qquad + \Pr(M(a^*)\ge t|\bX=\bx, U=1),
\end{align*}
where the second equality follows from (\ref{equ:ind3}), the third equality follows from (\ref{equ:seq2}) in Assumption \ref{ass:seq_ign}, and the last equality follows from the fact that $T(a)\ge t $ with probability one for any $M(a) = m\ge t$ given $U=1$.
Then, following (\ref{equ:seq1}) in Assumption \ref{ass:seq_ign} and Assumption \ref{ass:consistent}, the proceeding expression is equal to
\begin{align*}
& \int_0^t \frac{\Pr\{T(a)\ge t , M(a) = m| A = a, \bX=\bx, U=1\}}{\Pr\{M(a) = m| A = a, \bX=\bx, U=1\}}d F_{M(a^*)|A=a^*, \bX=\bx, U=1}(m)\\
&\qquad + \Pr(M(a^*)\ge t|A = a^*, \bX=\bx, U=1)\\
=& \int_0^t \Pr(T\ge t | M=m, A = a, \bX=\bx, U=1) d F_{M|A=a^*, \bX=\bx, U=1}(m)\\
&\qquad + \Pr(M > t|A = a^*, \bX=\bx, U=1).
\end{align*}
Then, the natural indirect and direct effects, as defined in equations (\ref{equ:NIE1}) and (\ref{equ:NDE1}), are equal to 
\begin{align*}
NIE_1(t;\bx) =& \int_0^t \Pr(T\ge t | M=m, A = 1, \bX=\bx, U=1)\\
&\qquad\times\left\{d F_{M|A=1, \bX=\bx, U=1}(m)-d F_{M|A=0, \bX=\bx, U=1}(m)\right\}\\
&\qquad + \Pr(M > t|A = 1, \bX=\bx, U=1) -\Pr(M > t|A = 0, \bX=\bx, U=1)
\end{align*}
and
\begin{align*}
NDE_1(t;\bx) =&\int_0^t \left\{\Pr(T\ge t | M=m, A = 1, \bX=\bx, U=1)\right.\\
&\qquad\left. -\Pr(T\ge t | M=m, A = 0, \bX=\bx, U=1)\right\} d F_{M|A=0, \bX=\bx, U=1}(m).
\end{align*}

For stratum with $U=2$, we have
\begin{align*}
 TE_2(t;\bx) =& \Pr(T(1)\ge t| A=1,\bX=\bx,U=2) - \Pr(T(0)\ge t| A = 0,\bX=\bx,U=2)\\
 =& \Pr(T\ge t| A=1,\bX=\bx,U=2) - \Pr(T\ge t| A = 0,\bX=\bx,U=2),
\end{align*}
where the two equalities follow from Assumptions \ref{ass:seq_ign} and \ref{ass:consistent}, respectively.
Similarly, for stratum with $U=3$, we have
\begin{align*}
 TE_3(t;\bx) =& \Pr(T(1)\ge t| A=1,\bX=\bx,U=3) - \Pr(T(0)\ge t| A = 0,\bX=\bx,U=3)\\
 =& \Pr(T\ge t| A=1,\bX=\bx,U=3) - \Pr(T\ge t| A = 0,\bX=\bx,U=3),
 % =& \Pr(T< t| A=0,\bX=\bx,U=3) - \Pr(T< t| A = 1,\bX=\bx,U=3),
\end{align*}

\subsection{Proof of Theorem \ref{thm:ident}} \label{append:ident}
We first consider the identification of stratum membership probabilities, as the first step to identify the stratum-specific effects.
By the definition of the strata, we have
\begin{align*}
 \Pr (M<\infty | A=0,\bX = \bx) =& \Pr (M(0)<\infty | A=0,\bX = \bx)\\
 =&\Pr (M(0)<\infty, M(1)<\infty | A=0,\bX = \bx)\\
 &\qquad + \Pr (M(0)<\infty,M(1) = \infty | A=0,\bX = \bx)\\
 =&\Pr (U=1 | A=0,\bX = \bx) + \Pr (U=2 | A=0,\bX = \bx)\\
 =&\Pr (U=1 | \bX = \bx) + \Pr (U=2 | \bX = \bx),
\end{align*}
where the last equality follows from Assumption \ref{ass:mem}.
Similarly, we have the following equalites:
\begin{align*}
 \Pr (M<\infty | A=1,\bX = \bx) =&\Pr (U=1 | \bX = \bx)\\
 \Pr (M=\infty | A=0,\bX = \bx) =&\Pr (U=3 | \bX = \bx)\\
 \Pr (M=\infty | A=1,\bX = \bx) =&\Pr (U=2 | \bX = \bx) + \Pr (U=3 | \bX = \bx).
\end{align*}
Therefore, the stratum membership probabilities $\Pr(U=u|\bX)$ can be identified by the observed data if there is no censoring and we can observed if $M=\infty$.
The identification of the quantities in the presence of censoring is discussed in the end of the section.

We further consider the distributions of the (observed) intermediate and primary event times to identify terms in the definition of stratum-specific effects.
First, consider the case if we observe $M=m<\infty$ for a subject assigned to treatment $A=1$.
Since it implies $M(1)<\infty$, this subject should have $U=1$ with probability one.
That is, for any $m<\infty$,
\begin{align*}
 \Pr(M=m|A=1,\bX = \bx) =& \Pr(M=m, M(1)\le T(1)|A=1,\bX = \bx)\\
 =& \Pr(M=m, M(0)\le T(0), M(1)\le T(1)|A=1,\bX = \bx)\\
 =& \Pr(M=m, U=1|A=1,\bX = \bx)\\
 =&\Pr(M=m|U=1, A=1,\bX = \bx)\Pr(U=1|A=1,\bX = \bx)\\
 =&\Pr(M(1)=m|\bX = \bx, U=1)\Pr(U=1|\bX = \bx),
\end{align*}
where the first equality follows from Assumption \ref{ass:consistent}, the second and third equalities follow from the definition of strata, and last equality follows from Assumption \ref{ass:mem}.
Similarly, we have for any $m\le t<\infty$,
\begin{align*}
 \Pr (T \ge t | M=m, A=1,\bX=\bx) = &\frac{\Pr (T < t , M=m| A=1,\bX=\bx)}{\Pr (M=m| A=1,\bX=\bx)}\\
 =&\frac{\Pr (T \ge t , M=m, M(1)\le T(1)| A=1,\bX=\bx)}{\Pr (M=m, M(1)\le T(1)| A=1,\bX=\bx)}\\
 =&\frac{\Pr (T \ge t , M=m, U=1| A=1,\bX=\bx)}{\Pr (M=m, U=1| A=1,\bX=\bx)}\\
 =&\Pr (T(1) \ge t| M(1)=m,\bX=\bx, U=1).
\end{align*}

For the case that we observe $M=m<\infty$ for a subject assigned to treatment $A=0$, the subject would possibly have $U=1$ or $U=2$, since both strata has $M(0)<\infty$.
Then, we have
\begin{align*}
 &\Pr(M=m|A=0,\bX = \bx)\\
 =& \Pr(M=m, M(0)\le T(0)|A=0,\bX = \bx)\\
 =& \Pr(M=m, M(0)\le T(0),M(1)\le T(1)|A=0,\bX = \bx)\\
 &\qquad + \Pr(M=m, M(0)\le T(0),M(1)=\infty|A=0,\bX = \bx)\\
 =& \Pr(M=m, U=1|A=0,\bX = \bx) + \Pr(M=m,U=2|A=0,\bX = \bx)\\
 =& \Pr(M=m |A=0,\bX = \bx, U=1)\Pr( U=1|\bX = \bx) \\
 &\qquad + \Pr(M=m|A=0,\bX = \bx,U=2)\Pr(U=2|\bX = \bx)\\
 =& \Pr(M(0)=m |\bX = \bx, U=1)\Pr( U=1|\bX = \bx) \\
 &\qquad + g_1(\Pr(M(0)=m |\bX = \bx, U=1);\bx)\Pr(U=2|\bX = \bx),
\end{align*}
where the last equality follows from Assumption \ref{ass:model},
and 
\begin{align*}
 &\Pr(T\ge t|M=m, A=0,\bX = \bx)\\
 =& \frac{\Pr (T \ge t , M=m| A=0,\bX=\bx)}{\Pr (M=m| A=0,\bX=\bx)}\\
 =& \frac{\Pr (T \ge t , M=m, M(0)\le T(0)| A=0,\bX=\bx)}{\Pr (M=m| A=0,\bX=\bx)}\\
 =& \frac{\sum_{u=1,2}\Pr (T(0)\ge t | M(0)=m,\bX=\bx, U = u)\Pr (M(0)=m|\bX=\bx, U = u)\Pr(U = u| \bX=\bx)}{\sum_{u=1,2}\Pr (M=m(0)|\bX=\bx,U=u))\Pr(U = u|\bX=\bx)}.
\end{align*}
Then, the natural indirect effect can be presented as
\begin{align}
NIE_1(t;\bx) =& \int_0^t \Pr(T\ge t | M=m, A = 1, \bX=\bx)\nonumber\\
&\qquad\times\left\{\Pr(M=m|M<\infty,A=1,\bX = \bx)-h_1^*(m;\bx)\right\}dm\nonumber\\
&\qquad + \Pr(M\le t|M<\infty,A=1,\bX = \bx) -\int_0^t h_1^*(m;\bx) dm. \label{equ:ident_gen_NIE1}
\end{align}
where $h_1^*(m;\bx)$ is the solution to the equation
\begin{align*}
& \Pr(M=m|A=0,\bX=\bx) = h_1(m;\bx) \Pr(M<\infty|A=1,\bX=\bx)\\
&\qquad+ g_1\{h_1(m;\bx);\bx\}\left\{\Pr(M<\infty|A=0,\bX=\bx)-\Pr(M<\infty|A=1,\bX=\bx)\right\}.
\end{align*}
Similarly, the natural direct effect can be presented as
\begin{equation}
NDE_1(t;\bx) =\int_0^t \left\{\Pr(T\ge t | M=m, A = 1, \bX=\bx) -h_2^*(m,t;\bx)\right\} h_1^*(m;\bx)dm,\label{equ:ident_gen_NDE1}
\end{equation}
where $h_2^*(m,t;\bx)$ is the solution to
\begin{align*}
&\Pr(T\ge t,M=m|A=0,\bX=\bx) = h_2(m,t;\bx) h_1^*(m;\bx)\Pr(M<\infty|A=1,\bX=\bx) \\
&\qquad+ g_2\{h_2(m,t;\bx);\bx\} g_1\{h_1^*(m;\bx);\bx\}\\
&\qquad \times\left\{\Pr(M<\infty|A=0,\bX=\bx)-\Pr(M<\infty|A=1,\bX=\bx)\right\}.
\end{align*}

By similar derivations, we have
\begin{align*}
 &\Pr(T\ge t, M = \infty | A=1,\bX = \bx)\\
 =&\Pr(T\ge t| A=1,\bX = \bx, U=2 )\Pr(U=2 |\bX = \bx)\\
 &\qquad + \Pr(T\ge t| A=1,\bX = \bx, U=3)\Pr(U=3 | \bX = \bx)\\
 =&g_3\left\{\Pr(T(1)\ge t|\bX = \bx, U=2 )\right\}\Pr(U=2 |\bX = \bx)\\
 &\qquad + \Pr(T(1)\ge t|\bX = \bx, U=3)\Pr(U=3 | \bX = \bx),
\end{align*}
and
\begin{align*}
 \Pr(T\ge t , M = \infty | A=0,\bX = \bx) 
 =&\Pr(T(0)\ge t | \bX = \bx, U=3)\Pr(U=3 | \bX = \bx).
\end{align*}
Then, the stratum-specific total effects for strata with $U=2$ and $U=3$ are
\begin{equation}
 TE_2(t;\bx) 
 = \int_0^t g_2 \{h_2^*(m,t;\bx);\bx\} g_1 \left\{h_1^*(m;\bx);\bx\right\}dm - g_3\left\{h_3^*(t;\bx);\bx\right\},\label{equ:ident_gen_TE2}
\end{equation}
and
\begin{equation}
 TE_3(t;\bx) = h_3^*(t;\bx) - \Pr(T\ge t | M = \infty, A=0,\bX = \bx),\label{equ:ident_gen:TE3}
\end{equation}
 where $h_3^*(t;\bx)$ is the solution to the equation
\begin{align*}
&\Pr(T=t,M<\infty|A=1,\bX=\bx) = h_3(t;\bx)\Pr (M=\infty | A=0,\bX = \bx) \\
&\qquad + g_3\left\{h_3(t;\bx);\bx\right\}\left\{\Pr (M=\infty | A=1,\bX = \bx)-\Pr (M=\infty | A=0,\bX = \bx)\right\}.
\end{align*}

In the special case that $g_k(\cdot;\bx)$ are identity functions, i.e., the stratum-specific joint distributions of $\{M(0),T0)\}$ are the same for strata $U=1$ and $U=2$ given $\bX=\bx$, and the stratum-specific distributions of $T(0)$ are the same for strata $U=2$ and $U=3$ given $\bX=\bx$,
the functions $h_k^*$'s have closed form
\begin{align*}
 &h_1^*(m;\bx) = \Pr(M=m|M<\infty,A=0,\bX = \bx),\\
 &h_2^*(m,t;\bx) =\Pr(T\ge t|M=m,A=0,\bX=\bx),\\
 &h_3^*(t;\bx) = \Pr(T<t |M = \infty, A=1,\bX = \bx).
\end{align*}
Then, the stratum-specific effects can be identified by
\begin{align*}
NIE_1(t;\bx) 
 =& \int_0^t \Pr (T \ge t | M=m, A=1,\bX=\bx)\left\{\Pr(M=m|M<\infty, A=1,\bX = \bx)\right.\\
&\qquad\left.-\Pr(M=m|M<\infty, A=0,\bX = \bx)\right\}dm\\
& + \Pr(M\le t|M<\infty,A=1,\bX = \bx) - \Pr(M\le t|M<\infty,A=0,\bX = \bx),\\
NDE_1(t;\bx)
=&\int_0^t \left\{\Pr(T\ge t|M=m,A=1,\bX=\bx) -\Pr (T \ge t | M=m, A=0,\bX=\bx)\right\}\\
&\qquad\times \Pr(M=m|M<\infty, A=0,\bX = \bx)dm,\\
TE_2(t;\bx) =& \int_0^t \Pr(T\ge t, M=m|M<\infty,A=1,\bX=\bx)dm\\
&\qquad - \Pr(T\ge t |M = \infty, A=0,\bX = \bx),
\end{align*}
and
\[TE_3(t;\bx)=\Pr(T\ge t |M = \infty, A=1,\bX = \bx) - \Pr(T\ge t |M = \infty, A=0,\bX = \bx).\]

In the presence of censoring, we cannot observe if $M=\infty$, such that previous formula cannot be directly applied.
However, we are still able to identify the quantities if we assume non-informative censoring and sufficient follow-up in strata (Assumption \ref{ass:cond_ind_cens}).
Particularly, we consider the marker process $I(M\le t)$ along with the event time $T$.
%Particularly, we consider the nonparametric estimation of survival function of $\max(M,T)$ in different $A$ and $\bX$.
Then, based on an extension of results in \cite{maller1992estimating}, the probability $\Pr(M=\infty|A,\bX)$ can be consistently estimated by the empirical value of the marker process at the last observed failure time.
By replacing terms related to $\Pr(M=\infty|A,\bX)$ by their estimators, we identify the stratum-specific mediation effects and total effects.

\section{Details on EM Algorithm}\label{append:EM}
Based on the likelihood function with known $U_i$, we are then able to propose an EM algorithm treating $U_i$ ($i=1, \dots, n$) as missing data.
In particular, the complete-data log-likelihood (with known $U_i$ for $i=1, \dots, n$) is given by 
\begin{align*}
&l_n(\bbeta,\bgamma, \balpha, \Lambda)\\
=&\sum_{i=1}^n\Bigg[I(U_i=1)\left\{\balpha_1^{\rm T}\tbX_i + \Delta_i^M\left(\log \Lambda_1\{Z_i\} + \bfeta_{M1}^{\rm T}\bW_i-e^{\bfeta_{R1}^{\rm T}\bW_i}\sum_{t_{2l}\le V_i}\lambda_{2l}\right)\right.\\
&\qquad\left. + \Delta_i^M\Delta_i^T\left(\log\Lambda_2\{V_i\} + \bfeta_{R1}^{\rm T}\bW_i\right) - \left(1-\Delta_i^T + \Delta_i^M\Delta_i^T\right)e^{\bfeta_{M1}^{\rm T}\bW_i}\sum_{t_{1l}\le Z_i}\lambda_{1l}\right\}\\
& + I(A_i=0, U_i=2)\left\{\balpha_2^{\rm T}\tbX_i + \Delta_i^M\left(\log \Lambda_1\{Z_i\} + \bfeta_{M2}^{\rm T}\tbX_i-e^{\bfeta_{R2}^{\rm T}\tbX_i}\sum_{t_{2l}\le V_i}\lambda_{2l}\right)\right.\\
& \qquad\left. + \Delta_i^M\Delta_i^T\left(\log \Lambda_2\{V_i\} + \bfeta_{R2}^{\rm T}\tbX_i\right)-\left(1-\Delta_i^T + \Delta_i^M\Delta_i^T\right)e^{\bfeta_{M2}^{\rm T}\tbX_i}\sum_{t_{1l}\le Z_i}\lambda_{1l}\right\}\\
& + I(A_i=1, U_i=2)(1-\Delta_i^M)\left\{\balpha_2^{\rm T}\tbX_i + \Delta_i^T\left(\log \Lambda_3\{Y_i\} + \bfeta_{T2}^{\rm T}\tbX_i\right)-e^{\bfeta_{T2}^{\rm T}\tbX_i}\sum_{t_{3l}\le Y_i}\lambda_{3l}\right\}\\
& + I(U_i=3)(1-\Delta_i^M)\left\{\Delta_i^T\left(\log \Lambda_3\{Y_i\} + \bfeta_{T3}^{\rm T}\bW_i\right)- e^{\bfeta_{T3}^{\rm T}\bW_i}\sum_{t_{3l}\le Y_i}\lambda_{3l}\right\}\\
&-\log \left\{1 + \exp\left(\balpha_1^{\rm T}\tbX_i\right) + \exp\left(\balpha_2^{\rm T}\tbX_i\right)\right\}\Bigg].
\end{align*}
In the E-step of the EM algorithm, we evaluate the conditional expectation $U_i$ for subjects $i=1, \dots, n$.
In particular, 
\begin{align*}
\widehat\Pr(U_i = 1) =&\Delta_i^M\left\{I(A_i=1) + I(A_i=0)\frac{B_{i1}}{B_{i1} + B_{i2}}\right\} + (1-\Delta_i^M)(1-\Delta_i^T)\frac{D_{i1}}{D_{i1} + D_{i2} + D_{i3}}\\
\widehat\Pr(U_i = 2) =&\Delta_i^MI(A_i=0)\frac{B_{i2}}{B_{i1} + B_{i2}} + (1-\Delta_i^M)\Delta_i^T I(A_i=1)\frac{C_{i2}}{C_{i2} + C_{i3}} \\
&\qquad + (1-\Delta_i^M)(1-\Delta_i^T)\frac{D_{i2}}{D_{i1} + D_{i2} + D_{i3}}\\
\widehat\Pr(U_i = 3) =&(1-\Delta_i^M)\Delta_i^T\left\{I(A_i=0) + I(A_i=1)\frac{C_{i3}}{C_{i2} + C_{i3}}\right\} \\
&\qquad + (1-\Delta_i^M)(1-\Delta_i^T)\frac{D_{i3}}{D_{i1} + D_{i2} + D_{i3}}
\end{align*}
where
\begin{align*}
B_{i1} =& \exp\left\{\balpha_1^{\rm T}\tbX_i + \bfeta_{M1}^{\rm T}\bW_i+ \Delta_i^T\left(\bfeta_{R1}^{\rm T}\bW_i\right)-e^{\bfeta_{M1}^{\rm T}\bW_i}\sum_{t_{1l}\le Z_i}\lambda_{1l}-e^{\bfeta_{R1}^{\rm T}\bW_i}\sum_{t_{2l}\le V_i}\lambda_{2l}\right\},\\
B_{i2} =& \exp\left\{\balpha_2^{\rm T}\tbX_i + \bfeta_{M2}^{\rm T}\tbX_i + \Delta_i^T\left(\bfeta_{R2}^{\rm T}\tbX_i\right)-e^{\bfeta_{M2}^{\rm T}\tbX_i}\sum_{t_{1l}\le Z_i}\lambda_{1l}-e^{\bfeta_{R2}^{\rm T}\tbX_i}\sum_{t_{2l}\le V_i}\lambda_{2l}\right\},\\
C_{i2} =& \exp\left(\balpha_2^{\rm T}\tbX_i + \bfeta_{T2}^{\rm T}\tbX_i-e^{\bfeta_{T2}^{\rm T}\tbX_i}\sum_{t_{3l}\le Y_i}\lambda_{3l}\right),\\
C_{i3} =& \exp\left(\bfeta_{T3}^{\rm T}\bW_i-e^{\bfeta_{T3}^{\rm T}\bW_i}\sum_{t_{3l}\le Y_i}\lambda_{3l}\right),\\
D_{i1} =&\exp\left(\balpha_1^{\rm T}\tbX_i -e^{\bfeta_{M1}^{\rm T}\bW_i}\sum_{t_{1l}\le Z_i}\lambda_{1l}\right),
\\
D_{i2} =&\exp\left(\balpha_2^{\rm T}\tbX_i\right)\left\{I(A_i=0)\exp\left(-e^{\bfeta_{M2}^{\rm T}\tbX_i}\sum_{t_{1l}\le Z_i}\lambda_{1l}\right)\right.\\
&\qquad\left. + I(A_i=1)\exp\left(-e^{\bfeta_{T2}^{\rm T}\tbX_i}\sum_{t_{3l}\le Y_i}\lambda_{3l}\right)\right\},
\end{align*}
and
\[ D_{i3} = \exp\left(-e^{\bfeta_{T3}^{\rm T}\bW_i}\sum_{t_{3l}\le Y_i}\lambda_{3l}\right).\]
In the M-step of the EM algorithm, we maximize the conditional expectation of the complete-data log-likelihood function.
In particular, we update $\Lambda_1$, $\Lambda_2$, and $\Lambda_3$ by
\begin{align*}
\lambda_{1l} =&\frac{\sum_{i=1}^n \Delta_i^MI(Z_i = t_{1l})}{\sum_{i=1}^n \left(1-\Delta_i^T + \Delta_i^M\Delta_i^T\right) I(Z_i \ge t_{1l})S_{i1}},\\
\lambda_{2l} =&\frac{\sum_{i=1}^n \Delta_i^M\Delta_i^T I(V_i = t_{2l})}{\sum_{i=1}^n\Delta_i^M I(V_i \ge t_{2l})S_{i2}}, \\
\lambda_{3l} =&\frac{\sum_{i=1}^n (1-\Delta_i^M)\Delta_i^T I(Y_i = t_{3l})}{\sum_{i=1}^n(1-\Delta_i^M) I(Y_i \ge t_{3l})S_{i3}},
\end{align*}
where 
\begin{align*}
S_{i1} =& \widehat\Pr(U_i=1) e^{\bfeta_{M1}^{\rm T}\bW_i} + \widehat\Pr(U_i=2)I(A_i=0) e^{\bfeta_{M2}^{\rm T}\tbX_i},\\
S_{i2} =& \widehat\Pr(U_i=1) e^{\bfeta_{R1}^{\rm T}\bW_i} + \widehat\Pr(U_i=2)I(A_i=0) e^{\bfeta_{R2}^{\rm T}\tbX_i},\\
S_{i3} =& \widehat\Pr(U_i=2)I(A_i=1) e^{\bfeta_{T2}^{\rm T}\tbX_i} + \widehat\Pr(U_i=3) e^{\bfeta_{T3}^{\rm T}\bW_i}.
\end{align*}
We update $\bfeta_{M1}$ by solving
\[\sum_{i=1}^n \Delta_i^M \left\{\widehat\Pr(U_i=1)\bW_i - \frac{\sum_{j=1}^nI(Z_j\ge Z_i)(1-\Delta_j^T + \Delta_j^M\Delta_j^T)\widehat\Pr(U_j=1) e^{\bfeta_{M1}^{\rm T}\bW_j}\bW_j}{\sum_{j=1}^nI(Z_j\ge Z_i)(1-\Delta_j^T + \Delta_j^M\Delta_j^T)S_{j1}}\right\} = \bzero,\]
and update $\bfeta_{M2}$ by solving
\begin{align*}
&\sum_{i=1}^n \Delta_i^M \left\{\widehat\Pr(U_i=2)I(A_i=0)\tbX_i \right.\\
&\qquad\left.- \frac{\sum_{j=1}^nI(Z_j\ge Z_i)(1-\Delta_j^T + \Delta_j^M\Delta_j^T)\widehat\Pr(U_j=2)I(A_j=0) e^{\beta_{M2} + \bgamma_{M2}^{\rm T}\bX_j}\tbX_j}{\sum_{j=1}^nI(Z_j\ge Z_i)(1-\Delta_j^T + \Delta_j^M\Delta_j^T)S_{j1}}\right\} = \bzero.
\end{align*}
We update $\bfeta_{R1}$ by solving
\begin{align*}
&\sum_{i=1}^n \Delta_i^M\Delta_i^T \left\{\widehat\Pr(U_i=1)\bW_i - \frac{\sum_{j=1}^nI(R_j\ge V_i)\Delta_j^M\widehat\Pr(U_j=1) e^{\bfeta_{R1}^{\rm T}\bW_j}\bW_j}{\sum_{j=1}^nI(R_j\ge V_i)\Delta_j^MS_{j2}}\right\} = \bzero, 
\end{align*}
and update $\bfeta_{R2}$ by solving
\begin{align*}
&\sum_{i=1}^n \Delta_i^M\Delta_i^T \left\{\widehat\Pr(U_i=2)I(A_i=0)\tbX_i \right.\\
&\qquad\left.- \frac{\sum_{j=1}^nI(R_j\ge V_i)\Delta_j^M\widehat\Pr(U_j=2)I(A_j=0) e^{\bfeta_{R2}^{\rm T}\tbX_j}\tbX_j}{\sum_{j=1}^nI(R_j\ge V_i)\Delta_j^MS_{j2}}\right\} = \bzero.
\end{align*}
We update $\bfeta_{T2}$ by solving
\begin{align*}
&\sum_{i=1}^n (1-\Delta_i^M)\Delta_i^T \left\{\widehat\Pr(U_i=2)I(A_i=1)\tbX_i \right.\\
&\qquad\left.- \frac{\sum_{j=1}^nI(Y_j \ge Y_i)(1-\Delta_j^M)\widehat\Pr(U_j=2)I(A_j=1) e^{\bfeta_{T2}^{\rm T}\tbX_j}\tbX_j}{\sum_{j=1}^nI(Y_j \ge Y_i)(1-\Delta_j^M)S_{j3}}\right\} = \bzero, 
\end{align*}
and update $\bfeta_{T3}$ by solving
\begin{align*}
&\sum_{i=1}^n (1-\Delta_i^M)\Delta_i^T \left\{\widehat\Pr(U_i=3)\bW_i \right.\\
&\qquad\left.- \frac{\sum_{j=1}^nI(Y_j \ge Y_i)(1-\Delta_j^M)\widehat\Pr(U_j=1) e^{\bfeta_{T3}^{\rm T}\bW_j}\bW_j}{\sum_{j=1}^nI(Y_j \ge Y_i)(1-\Delta_j^M)S_{j3}}\right\} = \bzero.
\end{align*}
Finally, we update $\balpha$ by solving
\begin{align*}
\sum_{i=1}^n\left\{\widehat\Pr(U_i = 1) - \frac{\exp\left(\balpha_1^{\rm T}\tbX_i\right)}{1 + \exp\left(\balpha_1^{\rm T}\tbX_i\right) + \exp\left(\balpha_2^{\rm T}\tbX_i\right)}\right\}\tbX_i = \bzero, \\
\sum_{i=1}^n\left\{\widehat\Pr(U_i = 2) - \frac{\exp\left(\balpha_2^{\rm T}\tbX_i\right)}{1 + \exp\left(\balpha_1^{\rm T}\tbX_i\right) + \exp\left(\balpha_2^{\rm T}\tbX_i\right)}\right\}\tbX_i = \bzero.
\end{align*}
Starting with $\btheta = \bzero $ and $\lambda_{kl} = 1/m_k$ for $k=1,2,3$, we iterate between the E-step and the M-step until convergence to obtain the nonparametric maximum likelihood estimators $(\wbtheta,\wcalA)$.

\section{Proofs of Asymptotic Results}
To prove the asymptotic results, we impose the following regularity conditions.
%Let $\|\cdot\|_{l^\infty[0,\tau]}$ be the supremum norm for functions on $[0,\tau]$, and $\|w\|_{BV[0,\tau]}$ the total variation of the function $w$ on $[0,\tau]$.
%Let $\mathP_n$ denote the empirical measure for $n$ independent subjects, $\mathP$ denote the true probability measure, and $\mathG_n = \sqrt n(\mathP_n-\mathP)$ denote the empirical process.
We introduce the following regularity conditions.
\begin{cond}\label{cond:1}
The function $\Lambda_{k0}$ is strictly increasing and continuously differentiable on $[0,\tau_k]$ for $k=1,2,3$.
The parameter $\wbtheta_0$ lies in the interior of a compact set in $\mathR^m$, where $m = 8(p+1)$ and $p$ is the dimension of $\bX$.
\end{cond}
\begin{cond}\label{cond:cov}
With probability 1, the covariate $\bX$ is bounded and not concentrated on a hyperplane of lower dimension.
\end{cond}
\begin{cond}\label{cond:3}
With probability 1, $\Pr(Z\ge \tau_1|\bX=\bx)>0$.
With probability 1, $\Pr(R\ge \tau_2|\bX=\bx,\Delta^M=1)>0$.
With probability 1, $\Pr(Y\ge \tau_3|\bX=\bx)>0$.
\end{cond}
\begin{cond}\label{cond:alpha} 
The true parameters satisfies $\bgamma_{M10}\ne \bgamma_{M20}$, $\bgamma_{R10}\ne \bgamma_{R20}$, and $\bgamma_{T20}\neq \bgamma_{T30}$.
\end{cond}
\begin{rem}
Conditions \ref{cond:1}-\ref{cond:3} are standard conditions for regression analysis of event time with censored data.
Condition \ref{cond:alpha} assumes that there are some differences in covariate effect in different strata, which is essential in distinguishing strata and model identification.
\end{rem}

The proof of Theorem 3  follows from the general theorems in \cite{zeng2010general}.
Particularly, using the notations in \cite{zeng2010general}, we write the likelihood function $L_n(\btheta,\calA)$ as
\[\prod_{i=1}^n\prod_{k=1}^K\prod_{t\le \tau_k} \Lambda_k\{t\}^{R_k(t)dN_k(t)}\Psi(\calO_i,\btheta,\calA),\]
where $R_{i1}(t) = I(Z_i\ge t)$, $R_{i2}(t) = \Delta_i^MI(V_i\ge t)$, $R_{i3}(t) = (1-\Delta_i^M)I(Y_i\ge t)$,
$N_{i1}(t) = \Delta_i^MI(Z_i\le t)$, $N_{i2}(t) = \Delta_i^M\Delta_i^TI(V_i\le t)$, $N_{i3}(t) = (1-\Delta_i^M)\Delta_i^TI(Y_i\le t)$,
and $\Psi(\calO_i,\btheta,\calA) = \Psi_{i1}(\btheta,\calA)^{\Delta_i^M}\left\{\Psi_{i2}(\btheta,\calA)^{\Delta_i^T}\Psi_{i3}(\btheta,\calA)^{1-\Delta_i^T}\right\}^{1-\Delta_i^M}$ with
\begin{align*}
\Psi_{i1}(\btheta,\calA)=&g_{i1}^{(1)}(\btheta,\calA)+g_{i1}^{(2)}(\btheta,\calA),\\
\Psi_{i2}(\btheta,\calA)=&g_{i2}^{(2)}(\btheta,\calA) + g_{i2}^{(3)}(\btheta,\calA),\\
\Psi_{i3}(\btheta,\calA)=&\Psi_{i2}(\btheta,\calA) + g_{i3}^{(1)}(\btheta,\calA) + g_{i3}^{(2)}(\btheta,\calA),\\
g_{i1}^{(1)}(\btheta,\calA)=&w_1(\bX_i;\balpha)e^{\bfeta_{M1}^{\rm T}\bW_i + \Delta_i^T\bfeta_{R1}^{\rm T}\bW_i}\exp\left\{-e^{\bfeta_{M1}^{\rm T}\bW_i}\Lambda_1(Z_i)-e^{\bfeta_{R1}^{\rm T}\bW_i}\Lambda_2(V_i)\right\},\\
g_{i1}^{(2)}(\btheta,\calA) =& I(A_i=0)w_2(\bX_i;\balpha)e^{\bfeta_{M2}^{\rm T}\tbX_i + \Delta_i^T\bfeta_{R2}^{\rm T}\tbX_i}\exp\left\{-e^{\bfeta_{M2}^{\rm T}\tbX_i}\Lambda_1(Z_i) -e^{\bfeta_{R2}^{\rm T}\tbX_i}\Lambda_2(V_i)\right\},\\
g_{i2}^{(2)}(\btheta,\calA)=&I(A_i=1)w_2(\bX_i;\balpha)e^{\Delta_i^T\bfeta_{T2}^{\rm T}\tbX_i}\exp\left\{-e^{\bfeta_{T2}^{\rm T}\tbX_i}\Lambda_3(Y_i)\right\}\\
g_{i2}^{(3)}(\btheta,\calA)=&w_3(\bX_i;\balpha)e^{\Delta_i^T\bfeta_{T3}^{\rm T}\bW_i}\exp\left\{-e^{\bfeta_{T3}^{\rm T}\bW_i}\Lambda_3(Y_i)\right\},\\
g_{i3}^{(1)}(\btheta,\calA)=&w_1(\bX_i;\balpha)\exp\left\{-e^{\bfeta_{M1}^{\rm T}\bW_i}\Lambda_1(Y_i)\right\}\\
g_{i3}^{(2)}(\btheta,\calA)=&I(A_i=0)w_2(\bX_i;\balpha)\exp\left\{-e^{\bfeta_{M2}^{\rm T}\tbX_i}\Lambda_1(Y_i)\right\}.
\end{align*}
Then, Theorem 3 follows if regularity conditions (C1) - (C7) in \cite{zeng2010general} holds.
Particularly, conditions (C1) and (C2) in \cite{zeng2010general} follow obviously from Conditions \ref{cond:1} - \ref{cond:3} and conditions (C3), (C4), and (C6) in \cite{zeng2010general} require the smoothness of the model structure, which can be easily verified by examining the structure of the model and likelihood.
Here, we only provide the detailed proofs of first and second identifiability conditions (conditions (C5) and (C7) in \cite{zeng2010general}).
\paragraph{First Identifiability Condition (C5)}
We consider the first identifiability condition (C5), i.e., we would like to prove if 
\[\prod_{k=1}^3\prod_{t\le \tau_k} \lambda_k^*(t)^{R_k(t)dN_k(t)}\Psi(\calO,\btheta^*,\calA^*) =\prod_{k=1}^3\prod_{t\le \tau_k} \lambda_{k0}(t)^{R_k(t)dN_k(t)}\Psi(\calO,\btheta_0,\calA_0)\]
almost surely, then $\btheta^*=\btheta_0$ and $\Lambda_k^*(t) = \Lambda_{k0}(t)$ for $t\in[0,\tau_k]$ for $k=1,2,3$.
We first assume $\Delta^M = \Delta^T = 1$ to find $\lambda_1^*(Z)\lambda_2^*(R)\Psi_1(\btheta^*,\calA^*) =\lambda_{10}(Z)\lambda_{20}(R) \Psi_1(\btheta_0,\calA_0)$.
We integrate $Z$ from $0$ to $z$ and $R$ from $0$ to $r$ in both sides to find that
\begin{align}
&w_1(\bX;\balpha^*)\exp\left\{-e^{\beta_{M1}^*A + {\bgamma_{M1}^*}^{\rm T}\bX}\Lambda_1^*(z)-e^{\beta_{R1}^*A + {\bgamma_{R1}^*}^{\rm T}\bX}\Lambda_2^*(r)\right\}\nonumber\\
&+ I(A=0)w_2(\bX;\balpha^*)\exp\left\{-e^{\beta_{M2}^* + {\bgamma_{M2}^*}^{\rm T}\bX}\Lambda_1^*(z)-e^{\beta_{R2}^* + {\bgamma_{R2}^*}^{\rm T}\bX}\Lambda_2^*(r)\right\}\nonumber\\
=&w_1(\bX;\balpha_0)\exp\left\{-e^{\beta_{M10}A + \bgamma_{M10}^{\rm T}\bX}\Lambda_{10}(z)-e^{\beta_{R10}A + \bgamma_{R10}^{\rm T}\bX}\Lambda_{20}(r)\right\}\nonumber\\
&+ I(A=0)w_2(\bX;\balpha_0)\exp\left\{-e^{\beta_{M20} + \bgamma_{M20}^{\rm T}\bX}\Lambda_{10}(z)-e^{\beta_{R20} + \bgamma_{R20}^{\rm T}\bX}\Lambda_{20}(r)\right\}\label{equ:ident1}
\end{align}
for any $z\in[0,\tau_1]$ and $r\in[0,\tau_2]$.
We first assume $z=r=0$ in equality and set $A=0,1$ to find
$w_1(\bX;\balpha^*) = w_1(\bX;\balpha_0)$ and $w_2(\bX;\balpha^*) = w_2(\bX;\balpha_0)$, such that $\balpha^*=\balpha_0$ followed from Condition \ref{cond:cov}.
We set $A=1$ and take algorithms of both sides of (\ref{equ:ident1}) to find 
\[-e^{\beta_{M1}^* + {\bgamma_{M1}^*}^{\rm T}\bX}\Lambda_1^*(z)-e^{\beta_{R1}^* + {\bgamma_{R1}^*}^{\rm T}\bX}\Lambda_2^*(r)=-e^{\beta_{M10} + \bgamma_{M10}^{\rm T}\bX}\Lambda_{10}(z)-e^{\beta_{R10} + \bgamma_{R10}^{\rm T}\bX}\Lambda_{20}(r)\]
for any $z\in[0,\tau_1]$ and $r\in[0,\tau_2]$, such that $\bgamma_{M1}^* = \bgamma_{M10}$, $\bgamma_{R1}^* = \bgamma_{R10}$, $e^{\beta_{M1}^*}\Lambda_1^*(z) = e^{\beta_{M10}}\Lambda_{10}(z)$ for $z\in[0,\tau_1]$, and $e^{\beta_{R1}^*}\Lambda_2^*(r) = e^{\beta_{R10}}\Lambda_{20}(r)$ for $r\in[0,\tau_2]$.
We then set $A=0$ in equality (\ref{equ:ident1}) to find
\begin{align*}
&w_1(\bX;\balpha_0)\exp\left\{-e^{\beta_{M10}-\beta_{M1}^* + \bgamma_{M10}^{\rm T}\bX}\Lambda_{10}(z)-e^{\beta_{R10}-\beta_{R1}^* + \bgamma_{R10}^{\rm T}\bX}\Lambda_{20}(r)\right\}\\
&+ w_2(\bX;\balpha_0)\exp\left\{-e^{\beta_{M10}-\beta_{M1}^* + \beta_{M2}^* + {\bgamma_{M2}^*}^{\rm T}\bX}\Lambda_{10}(z)-e^{\beta_{R10}-\beta_{R1}^* + \beta_{R2}^* + {\bgamma_{R2}^*}^{\rm T}\bX}\Lambda_{20}(r)\right\}\\
=&w_1(\bX;\balpha_0)\exp\left\{-e^{ \bgamma_{M10}^{\rm T}\bX}\Lambda_{10}(z)-e^{\bgamma_{R10}^{\rm T}\bX}\Lambda_{20}(r)\right\}\\
&+ w_2(\bX;\balpha_0)\exp\left\{-e^{\beta_{M20} + \bgamma_{M20}^{\rm T}\bX}\Lambda_{10}(z)-e^{\beta_{R20} + \bgamma_{R20}^{\rm T}\bX}\Lambda_{20}(r)\right\}
\end{align*}
for any $z\in[0,\tau_1]$ and $r\in[0,\tau_2]$.
Since $\bgamma_{M10}\ne \bgamma_{M20}$ and $\bgamma_{R10}\ne \bgamma_{R20}$ by Condition \ref{cond:alpha}, we have $\beta_{M1}^* = \beta_{M10}$, $\beta_{R1}^* = \beta_{R10}$, $\beta_{M2}^* = \beta_{M20}$, $\beta_{R2}^* = \beta_{R20}$, $\bgamma_{M2}^* = \bgamma_{M20}$, and $\bgamma_{R2}^* = \bgamma_{R20}$. 

Similarly, we may assume $\Delta^M=0$ and $\Delta^T=1$ to find $\lambda_3(Y)^*\Psi_2(\btheta^*,\calA^*) = \lambda_{30}(Y)\Psi_2(\btheta_0,\calA_0)$.
We integrate $Y$ from 0 to $y$ in both sides of the resulting equation to find that
\begin{align}
 &I(A=1)w_2(\bX;\balpha_0)\exp\left(-e^{\beta_{T2}^* + {\bgamma_{T2}^*}^{\rm T}\bX}\Lambda_3^*(y)\right) + w_3(\bX;\balpha_0)\exp\left(-e^{\beta_{T3}^*A + {\bgamma_{T3}^*}^{\rm T}\bX}\Lambda_3^*(y)\right)\nonumber\\
 &=I(A=1)w_2(\bX;\balpha_0)\exp\left(-e^{\beta_{T20} + \bgamma_{T20}^{\rm T}\bX}\Lambda_{30}(y)\right) + w_3(\bX;\balpha_0)\exp\left(-e^{\beta_{T30}A + \bgamma_{T30}^{\rm T}\bX}\Lambda_{30}(y)\right).\label{equ:ident3}
\end{align}
We set $A=0$ and take logarithms of both sides of equation (\ref{equ:ident3}) to find that 
\[e^{{\bgamma_{T3}^*}^{\rm T}\bX}\Lambda_3^*(y) = e^{ \bgamma_{T30}^{\rm T}\bX}\Lambda_{30}(y),\]
such that $\bgamma_{T3}^* = \bgamma_{T30}$ and $\Lambda_3^*(y) = \Lambda_{30}(y)$ for $y\in[0,\tau_3]$.
We then set $A=1$ in equality (\ref{equ:ident3}) to find
\begin{align*}
 & w_2(\bX;\balpha_0)\exp\left(-e^{\beta_{T2}^* + {\bgamma_{T2}^*}^{\rm T}\bX}\Lambda_{30}(y)\right)+ w_3(\bX;\balpha_0)\exp\left(-e^{\beta_{T3}^* + \bgamma_{T30}^{\rm T}\bX}\Lambda_{30}(y)\right)\\
 =& w_2(\bX;\balpha_0)\exp\left(-e^{\beta_{T20} + \bgamma_{T20}^{\rm T}\bX}\Lambda_{30}(y)\right)+ w_3(\bX;\balpha_0)\exp\left(-e^{\beta_{T30} + \bgamma_{T30}^{\rm T}\bX}\Lambda_{30}(y)\right)
\end{align*}
for any $y\in[0,\tau_3]$.
By Condition \ref{cond:alpha}, $\bgamma_{T20}\ne \bgamma_{T30}$.
Therefore, $\beta_{T2}^* = \beta_{T20}$, $\beta_{T3}^* = \beta_{T30}$, and $\bgamma_{T2}^* = \bgamma_{T20}$.
The identifiability condition (C5) in \cite{zeng2010general} then holds.

\paragraph{Second Identifiability Condition (C7)}
We verify the second identifiability condition (C7) in \cite{zeng2010general}, i.e., if with probability one,
\begin{equation}
\sum_{k=1}^3\int h_k(t)R_k(t)dN_k(t) + \frac{\dot\Psi_\btheta(\calO;\btheta_0,\calA_0)^{\rm T}\bv + \sum_{k=1}^3\dot\Psi_k(\calO;\btheta_0,\calA_0)[\int h_kd\Lambda_{k0}]}{\Psi(\calO;\btheta_0,\calA_0)}=0,\label{equ:infor}
\end{equation}
for some constant vector $\bv\in\mathR^m$ and $h_k\in BV[0,\tau_k]$ $(k=1,2,3)$, then $\bv=\bzero$ and $h_k=0$ for $k=1,2,3$.

Let $\bv = (\bv_{M1}^{\rm T}, \bv_{R1}^{\rm T}, \bv_{M2}^{\rm T}, \bv_{R2}^{\rm T}, \bv_{T2}^{\rm T}, \bv_{T3}^{\rm T}, \bv_{\balpha_1}^{\rm T}, \bv_{\balpha_2}^{\rm T})^{\rm T}$.
Write $\bW = (A,\bX^{\rm T})^{\rm T}$, $\bX_0 = (0,\bX^{\rm T})^{\rm T}$, and note that $\tbX = (1,\bX^{\rm T})^{\rm T}$.
Then,
\[ \frac{\dot\Psi_\btheta(\calO;\btheta_0,\calA_0)^{\rm T}\bv }{\Psi(\calO;\btheta_0,\calA_0)} = \dot\Psi_{\bfeta}(\bv) + \dot\Psi_{\balpha}(\bv),\]
where 
\begin{align*}
 \dot\Psi_{\bfeta}(\bv) =&\Delta^M\frac{g_1^{(1)}(\btheta_0,\calA_0)\bW^{\rm T}\bv_{M1}+g_1^{(2)}(\btheta_0,\calA_0)\tbX^{\rm T}\bv_{M2}}{\Psi_1(\btheta_0,\calA_0)}\\
 &-\Delta^M\frac{g_1^{(1)}(\btheta_0,\calA_0) e^{\bfeta_{M10}^{\rm T}\bW}\bW^{\rm T}\bv_{M1} + g_1^{(2)}(\btheta_0,\calA_0) e^{\bfeta_{M20}^{\rm T}\tbX}\tbX^{\rm T}\bv_{M2}}{\Psi_1(\btheta_0,\calA_0)} \Lambda_{10}(Z)\\
 & -(1-\Delta^M)(1-\Delta^T)\frac{g_3^{(1)}(\btheta_0,\calA_0) e^{\bfeta_{M10}^{\rm T}\bW}\bW^{\rm T}\bv_{M1}+g_3^{(2)}(\btheta_0,\calA_0) e^{\bfeta_{M20}^{\rm T}\tbX}\tbX^{\rm T}\bv_{M2}}{\Psi_3(\btheta_0,\calA_0)} \Lambda_{10}(Y)\\
 & + \Delta^M\Delta^T\frac{g_1^{(1)}(\btheta_0,\calA_0)\bW^{\rm T}\bv_{R1} + g_1^{(2)}(\btheta_0,\calA_0)\tbX^{\rm T}\bv_{R2}}{\Psi_1(\btheta_0,\calA_0)}\\
 & - \Delta^M\frac{g_1^{(1)}(\btheta_0,\calA_0) e^{\bfeta_{R10}^{\rm T}\bW}\bW^{\rm T}\bv_{R1} + g_1^{(2)}(\btheta_0,\calA_0) e^{\bfeta_{R20}^{\rm T}\tbX}\tbX^{\rm T}\bv_{R2}}{\Psi_1(\btheta_0,\calA_0)}\Lambda_{20}(R)\\
 &+ (1-\Delta^M)\Delta^T\frac{g_2^{(2)}(\btheta_0,\calA_0)\tbX^{\rm T}\bv_{T2} + g_2^{(3)}(\btheta_0,\calA_0)\bW^{\rm T}\bv_{T3}}{\Psi_2(\btheta_0,\calA_0)}\\
 &-\left\{ \frac{(1-\Delta^M)\Delta^T}{\Psi_2(\btheta_0,\calA_0)} + \frac{(1-\Delta^M)(1-\Delta^T)}{\Psi_3(\btheta_0,\calA_0)} \right\}\\
 &\qquad\times\left\{g_2^{(2)}(\btheta_0,\calA_0) e^{\bfeta_{T20}^{\rm T}\tbX}\tbX^{\rm T}\bv_{T2} + g_2^{(3)}(\btheta_0,\calA_0) e^{\bfeta_{T30}^{\rm T}\bW}\bW^{\rm T}\bv_{T3}\right\} \Lambda_{30}(Y),
 \end{align*}
 and
 \begin{align*}
 \dot\Psi_{\balpha}(\bv) = & \left[\left\{ \Delta^M\frac{g_1^{(1)}(\btheta_0,\calA_0)}{\Psi_1(\btheta_0,\calA_0)}+ (1-\Delta^M)(1-\Delta^T)\frac{g_3^{(1)}(\btheta_0,\calA_0)}{\Psi_3(\btheta_0,\calA_0)}\right\} \left\{w_2(\bX;\balpha_0)+w_3(\bX;\balpha_0)\right\} \right.\\
 &\qquad\left. - \left\{\Delta^M\frac{g_1^{(2)}(\btheta_0,\calA_0)} {\Psi_1(\btheta_0,\calA_0)} + (1-\Delta^M)\Delta^T\frac{g_2^{(2)}(\btheta_0,\calA_0) +g_2^{(3)}(\btheta_0,\calA_0)}{\Psi_2(\btheta_0,\calA_0)} \right.\right.\\
 &\qquad\left.\left. + (1-\Delta^M)(1-\Delta^T)\frac{g_2^{(2)}(\btheta_0,\calA_0) + g_2^{(3)}(\btheta_0,\calA_0) + g_3^{(2)}(\btheta_0,\calA_0)}{\Psi_3(\btheta_0,\calA_0)} \right\}w_1(\bX;\balpha_0) \right]\tbX^{\rm T}\bv_{\balpha1}\\
&+\left[\left\{\Delta^M\frac{g_1^{(2)}(\btheta_0,\calA_0)} {\Psi_1(\btheta_0,\calA_0)} + (1-\Delta^M)\Delta^T\frac{g_2^{(2)}(\btheta_0,\calA_0)}{\Psi_2(\btheta_0,\calA_0)}\right.\right.\\
&\qquad\left.\left. +(1-\Delta^M)(1-\Delta^T)\frac{g_2^{(2)}(\btheta_0,\calA_0)+g_3^{(2)}(\btheta_0,\calA_0)}{\Psi_3(\btheta_0,\calA_0)} \right\}\left\{w_1(\bX;\balpha_0)+w_3(\bX;\balpha_0)\right\}\right.\\
&\qquad\left.-\left\{\Delta^M\frac{g_1^{(1)}(\btheta_0,\calA_0)}{\Psi_1(\btheta_0,\calA_0)} + (1-\Delta^M)\Delta^T\frac{g_2^{(3)}(\btheta_0,\calA_0)}{\Psi_2(\btheta_0,\calA_0)} \right.\right.\\
&\qquad \left.\left.+ (1-\Delta^M)(1-\Delta^T)\frac{g_2^{(3)}(\btheta_0,\calA_0) + g_3^{(1)}(\btheta_0,\calA_0)}{\Psi_3(\btheta_0,\calA_0)} \right\} w_2(\bX;\balpha_0)\right]\tbX^{\rm T}\bv_{\balpha2}.
\end{align*}
In addition, we have 
\begin{align*}
&\sum_{k=1}^K\frac{\dot\Psi_k(\calO;\btheta_0,\calA_0)[\int h_kd\Lambda_{k0}]}{\Psi(\calO;\btheta_0,\calA_0)} \\
=& - \int \left[\Delta^M\frac{g_1^{(1)}(\btheta_0,\calA_0) e^{\bfeta_{M10}^{\rm T}\bW} +g_1^{(2)}(\btheta_0,\calA_0) e^{\bfeta_{M20}^{\rm T}\tbX}}{\Psi_1(\btheta_0,\calA_0)} I(Z\ge t)\right.\\
&\qquad\left.+(1-\Delta^M)(1-\Delta^T)\frac{g_3^{(1)}(\btheta_0,\calA_0) e^{\bfeta_{M10}^{\rm T}\bW} + g_3^{(2)}(\btheta_0,\calA_0) e^{\bfeta_{M20}^{\rm T}\tbX}}{\Psi_3(\btheta_0,\calA_0)} I(Y\ge t)\right] h_1(t)d\Lambda_{10}(t)\\
& - \int \Delta^M\frac{g_1^{(1)}(\btheta_0,\calA_0) e^{\bfeta_{R10}^{\rm T}\bW} +g_1^{(2)}(\btheta_0,\calA_0) e^{\bfeta_{R20}^{\rm T}\tbX}}{\Psi_1(\btheta_0,\calA_0)} I(R\ge t)h_2(t)d\Lambda_{20}(t)\\
& -\int\left\{\frac{(1-\Delta^M)\Delta^T}{\Psi_2(\btheta_0,\calA_0)}+\frac{(1-\Delta^M)(1-\Delta^T)}{\Psi_3(\btheta_0,\calA_0)}\right\} \left\{ g_2^{(2)}(\btheta_0,\calA_0) e^{\bfeta_{T20}^{\rm T}\tbX} + g_2^{(3)}(\btheta_0,\calA_0) e^{\bfeta_{T30}^{\rm T}\bW}\right\}\\
&\qquad \times I(Y\ge t)h_3(t)d\Lambda_{30}(t).
\end{align*}

Since equality (\ref{equ:infor}) holds with probability one, we may arbitrarily set the values of the observed data in the domain such that the equality still holds.
We first set $\Delta^M=0$ and $\Delta^T=1$ to find 
\begin{align}
&h_3(T) +\frac{ g_2^{(2)}(\btheta_0,\calA_0)\tbX^{\rm T}\bv_{T2} + g_2^{(3)}(\btheta_0,\calA_0)\bW^{\rm T}\bv_{T3} }{\Psi_2(\btheta_0,\calA_0)}\nonumber\\
&\qquad- \frac{g_2^{(2)}(\btheta_0,\calA_0) e^{\bfeta_{T20}^{\rm T}\tbX}\tbX^{\rm T}\bv_{T2} + g_2^{(3)}(\btheta_0,\calA_0) e^{\bfeta_{T30}^{\rm T}\bW}\bW^{\rm T}\bv_{T3}}{\Psi_2(\btheta_0,\calA_0)}\Lambda_{30}(T)\nonumber\\
&\qquad - \frac{g_2^{(2)}(\btheta_0,\calA_0) +g_2^{(3)}(\btheta_0,\calA_0)}{\Psi_2(\btheta_0,\calA_0)} w_1(\bX;\balpha_0) \tbX^{\rm T}\bv_{\balpha1}\nonumber\\
&\qquad+\left(\frac{g_2^{(2)}(\btheta_0,\calA_0)}{\Psi_2(\btheta_0,\calA_0)} \left\{w_1(\bX;\balpha_0)+w_3(\bX;\balpha_0)\right\} -\frac{g_2^{(3)}(\btheta_0,\calA_0)}{\Psi_2(\btheta_0,\calA_0)}w_2(\bX;\balpha_0)\right)\tbX^{\rm T}\bv_{\balpha2}\nonumber\\
&\qquad-\int \frac{g_2^{(2)}(\btheta_0,\calA_0) e^{\bfeta_{T20}^{\rm T}\tbX} + g_2^{(3)}(\btheta_0,\calA_0) e^{\bfeta_{T30}^{\rm T}\bW}}{\Psi_2(\btheta_0,\calA_0)} I(T\ge t)h_3(t)d\Lambda_{30}(t)=0.\label{equ:M0T1}
\end{align}
We set $A=0$ to find 
\begin{align*}
&h_3(T) +\bX_0^{\rm T}\bv_{T3}-\bX_0^{\rm T}\bv_{T3} e^{\bfeta_{T30}^{\rm T}\bX_0} \Lambda_{30}(T) - w_1(\bX;\balpha_0) \tbX^{\rm T}\bv_{\balpha1} \\
&\qquad -w_2(\bX;\balpha_0)\tbX^{\rm T}\bv_{\balpha2} -\int e^{\bfeta_{T30}^{\rm T}\bX_0} I(T\ge t)h_3(t)d\Lambda_{30}(t)=0.
\end{align*}
We multiple both sides by $\lambda_{30}(T) \exp(\bfeta_{T30}^{\rm T}\bX_0)\exp\left\{-\Lambda_{30}(T) \exp(\bfeta_{T30}^{\rm T}\bX_0)\right\}$ and integrate $T$ from $0$ to $t$ for some $t\in[0,\tau_3]$ to obtain 
\begin{align*}
&\int_0^t e^{\bfeta_{T30}^{\rm T}\bX_0} \left\{ h_3(s) + \bX_0^{\rm T}\bv_{T3}\right\} d\Lambda_{30}(s) \exp\left\{-\Lambda_{30}(t) e^{\bfeta_{T30}^{\rm T}\bX_0} \right\} \\
&\qquad - w_1(\bX;\balpha_0) \tbX^{\rm T}\bv_{\balpha1} -w_2(\bX;\balpha_0)\tbX^{\rm T}\bv_{\balpha2}=0.
\end{align*}
We differentiate both sides with respect to $t$ to find
\begin{align*}
& h_3(t) +\bX_0^{\rm T}\bv_{T3} +\int_0^t \left\{h_3(s) + \bX_0^{\rm T}\bv_{T3}\right\} e^{\bfeta_{T30}^{\rm T}\bX_0}d\Lambda_{30}(s) =0.
\end{align*}
This is a homogeneous integral equation for $h_3(t) +\bX_0^{\rm T}\bv_{T3}$ and has zero solution.
That is, $h_3(t) +\bX_0^{\rm T}\bv_{T3}=0$.
It then follows from Condition \ref{cond:cov} that the last $p$ elemets of $\bv_{T3}$ are zeros and $h_3(t)=0$ for $t\in[0,\tau_3]$.
Then, it follows from Conditions \ref{cond:cov} and \ref{cond:alpha} that $\bv_{\balpha1} = \bv_{\balpha2}=\bzero$.
Let $v_{T3,1}$ denote the first element of $\bv_{T3}$.
Then, equality (\ref{equ:M0T1}) with $A=1$ can be reduced to
\begin{align}
& w_2(\bX;\balpha_0)e^{\bfeta_{T20}^{\rm T}\tbX}\exp\left\{-e^{\bfeta_{T20}^{\rm T}\tbX}\Lambda_{30}(t)\right\}\tbX^{\rm T}\bv_{T2}\left\{1- e^{\bfeta_{T20}^{\rm T}\tbX}\Lambda_{30}(t)\right\}\nonumber \\
&\qquad + w_3(\bX;\balpha_0)e^{\bfeta_{T30}^{\rm T}\tbX}\exp\left\{-e^{\bfeta_{T30}^{\rm T}\tbX}\Lambda_{30}(t)\right\} v_{T3,1}\left\{1- e^{\bfeta_{T30}^{\rm T}\tbX}\Lambda_{30}(t)\right\} =0.\label{equ:invert_1}
\end{align}
for any $t\in[0,\tau_3]$.

Similarly, we set $\Delta^M = \Delta^T=1$ in equality (\ref{equ:infor}) to find 
\begin{align}
& h_1(Z) + h_2(R) +\frac{ g_1^{(1)}(\btheta_0,\calA_0)\bW^{\rm T}\bv_{M1}+g_1^{(2)}(\btheta_0,\calA_0)\tbX^{\rm T}\bv_{M2} }{\Psi_1(\btheta_0,\calA_0)}\nonumber\\
 &\qquad + \frac{ g_1^{(1)}(\btheta_0,\calA_0)\bW^{\rm T}\bv_{R1} + g_1^{(2)}(\btheta_0,\calA_0)\tbX^{\rm T}\bv_{R2} }{\Psi_1(\btheta_0,\calA_0)}\nonumber\\
&\qquad - \frac{g_1^{(1)}(\btheta_0,\calA_0) e^{\bfeta_{M10}^{\rm T}\bW}\bW^{\rm T}\bv_{M1} + g_1^{(2)}(\btheta_0,\calA_0) e^{\bfeta_{M20}^{\rm T}\tbX}\tbX^{\rm T}\bv_{M2}}{\Psi_1(\btheta_0,\calA_0)} \Lambda_{10}(Z)\nonumber\\
 &\qquad- \frac{g_1^{(1)}(\btheta_0,\calA_0) e^{\bfeta_{R10}^{\rm T}\bW}\bW^{\rm T}\bv_{R1} + g_1^{(2)}(\btheta_0,\calA_0) e^{\bfeta_{R20}^{\rm T}\tbX}\tbX^{\rm T}\bv_{R2}}{\Psi_1(\btheta_0,\calA_0)} \Lambda_{20}(R)\nonumber\\
& - \int \frac{g_1^{(1)}(\btheta_0,\calA_0) e^{\bfeta_{R10}^{\rm T}\bW} +g_1^{(2)}(\btheta_0,\calA_0) e^{\bfeta_{R20}^{\rm T}\tbX}}{\Psi_1(\btheta_0,\calA_0)} I(R\ge t)h_2(t)d\Lambda_{20}(t) =0\label{equ:M1T1}.
\end{align}
We first set $A=1$ in equality (\ref{equ:M1T1}) to find 
\begin{align*}
& h_1(Z) + h_2(R) +\tbX^{\rm T}\bv_{M1} + \tbX^{\rm T}\bv_{R1}\\
&\qquad -\int e^{\bfeta_{M10}^{\rm T}\tbX}\tbX^{\rm T}\bv_{M1}I(Z\ge t) d\Lambda_{10}(t) -\int e^{\bfeta_{R10}^{\rm T}\tbX}\tbX^{\rm T}\bv_{R1} I(R\ge t)d\Lambda_{20}(t)\\
 &\qquad - \int e^{\bfeta_{M10}^{\rm T}\tbX} I(Z\ge t) h_1(t)d\Lambda_{10}(t) - \int e^{\bfeta_{R10}^{\rm T}\tbX} I(R\ge t)h_2(t)d\Lambda_{20}(t) =0.
\end{align*}
We multiple both sides by $\lambda_{10}(Z) \exp(\bfeta_{M10}^{\rm T}\tbX)\exp\left\{-\Lambda_{10}(Z) \exp(\bfeta_{M10}^{\rm T}\tbX)\right\}$ and integrate $Z$ from $0$ to $t_1$ for some $t_1\in[0,\tau_1]$ to obtain
\begin{align*}
& \int_0^{t_1} e^{\bfeta_{M10}^{\rm T}\tbX} \left\{ h_1(s_1) + \tbX^{\rm T}\bv_{M1}\right\} d\Lambda_{10}(s_1) \exp\left\{-\Lambda_{10}(t_1) e^{\bfeta_{M10}^{\rm T}\tbX} \right\} + h_2(R) + \tbX^{\rm T}\bv_{R1}\\
&\qquad -\int e^{\bfeta_{R10}^{\rm T}\tbX}\tbX^{\rm T}\bv_{R1} I(R\ge t)d\Lambda_{20}(t) - \int e^{\bfeta_{R10}^{\rm T}\tbX} I(R\ge t)h_2(t)d\Lambda_{20}(t) =0.
\end{align*}
Then, we multiple both sides by $\lambda_{20}(R) \exp(\bfeta_{R10}^{\rm T}\tbX)\exp\left\{-\Lambda_{20}(R) \exp(\bfeta_{R10}^{\rm T}\tbX)\right\}$ and integrate $R$ from $0$ to $t_2$ for some $t_2\in[0,\tau_2]$ to obtain 
\begin{align*}
& \int_0^{t_1} e^{\bfeta_{M10}^{\rm T}\tbX} \left\{ h_1(s_1) + \tbX^{\rm T}\bv_{M1}\right\} d\Lambda_{10}(s_1) \exp\left\{-\Lambda_{10}(t_1) e^{\bfeta_{M10}^{\rm T}\tbX} \right\} \\
&\qquad+ \int_0^{t_2} e^{\bfeta_{R10}^{\rm T}\tbX} \left\{ h_2(s_2) + \tbX^{\rm T}\bv_{R1}\right\} d\Lambda_{20}(s_2) \exp\left\{-\Lambda_{20}(t_2) e^{\bfeta_{R10}^{\rm T}\tbX} \right\} =0.
\end{align*}
Since this equality holds for any $t_1\in[0,\tau_1]$ and $t_2\in[0,\tau_2]$, the two terms in the left hand side are zero with probability one.
Again, since the two terms are homogeneous integral equations and have zero solutions, we have $h_1(t_1) +\tbX^{\rm T}\bv_{M1}=0$ for $t_1\in[0,\tau_1]$ and $h_2(t_2) +\tbX^{\rm T}\bv_{R1}=0$ for $t_2\in[0,\tau_2]$ .
It then follows from Condition \ref{cond:cov} that the last $p$ elemets of $\bv_{M1}$ and the last $p$ elemets of $\bv_{R1}$ are zeros, $h_1(t_1)=-v_{M1,1}$ for $t_1\in[0,\tau_1]$, and $h_2(t_2)=-v_{R1,1}$ for $t_2\in[0,\tau_2]$, where $v_{M1,1}$ and $v_{R1,1}$ are the first elements of $\bv_{M1}$ and $\bv_{R1}$, respectively.
We replace the terms and set $A=0$ in equality (\ref{equ:M1T1}) to find 
\begin{align}
& - w_1(\bX;\balpha_0)e^{\bfeta_{M10}^{\rm T}\bX_0}\exp\left\{-e^{\bfeta_{M10}^{\rm T}\bX_0}\Lambda_{10}(t_1)\right\}e^{ \bfeta_{R10}^{\rm T}\bX_0}\exp\left\{-e^{\bfeta_{R10}^{\rm T}\bX_0}\Lambda_{20}(t_2)\right\}\nonumber\\
&\qquad\times \left[v_{M1,1}\left\{1-e^{\bfeta_{M10}^{\rm T}\bX_0} \Lambda_{10}(t_1)\right\} + v_{R1,1}\left\{1- e^{\bfeta_{R10}^{\rm T}\bX_0} \Lambda_{20}(t_2)\right\}\right]\nonumber\\
& + w_2(\bX;\balpha_0)e^{\bfeta_{M20}^{\rm T}\tbX}\exp\left\{-e^{\bfeta_{M20}^{\rm T}\tbX}\Lambda_{10}(t_1)\right\}e^{ \bfeta_{R20}^{\rm T}\tbX}\exp\left\{-e^{\bfeta_{R20}^{\rm T}\tbX}\Lambda_{20}(t_2)\right\}\nonumber\\
&\qquad\times \left[\left(\tbX^{\rm T}\bv_{M2}-v_{M1,1}\right)\left\{1- e^{\bfeta_{M20}^{\rm T}\tbX}\Lambda_{10}(t_1) \right\} + \left(\tbX^{\rm T}\bv_{R2} - v_{R1,1}\right)\left\{1- e^{\bfeta_{R20}^{\rm T}\tbX} \Lambda_{20}(t_2) \right\} \right] =0\label{equ:invert_2}
\end{align}
for any $t_1\in[0,\tau_1]$ and $t_2\in[0,\tau_2]$.

Then, we set $\Delta^M = 1$ and $\Delta^T=0$ in equality (\ref{equ:infor}) to find 
\begin{align*}
&g_1^{(1)}(\btheta_0,\calA_0)Av_{M1,1}+g_1^{(2)}(\btheta_0,\calA_0)\tbX^{\rm T}\bv_{M2}\nonumber\\
& + g_1^{(1)}(\btheta_0,\calA_0) e^{\bfeta_{M10}^{\rm T}\bW}\left(v_{M1,1}-Av_{M1,1}\right) \Lambda_{10}(Z)+g_1^{(2)}(\btheta_0,\calA_0) e^{\bfeta_{M20}^{\rm T}\tbX}\left(v_{M1,1} - \tbX^{\rm T}\bv_{M2}\right)\Lambda_{10}(Z)\\
& + g_1^{(1)}(\btheta_0,\calA_0) e^{\bfeta_{R10}^{\rm T}\bW}\left(v_{R1,1}-Av_{R1,1}\right)\Lambda_{20}(R) +g_1^{(2)}(\btheta_0,\calA_0) e^{\bfeta_{R20}^{\rm T}\tbX}\left(v_{R1,1}- \tbX^{\rm T}\bv_{R2}\right)\Lambda_{20}(R) =0.
\end{align*}
We then set $A=0$ to obtain 
\begin{align*} 
& g_1^{(1)}(\btheta_0,\calA_0) e^{\bfeta_{M10}^{\rm T}\bW} v_{M1,1} \Lambda_{10}(Z) + g_1^{(1)}(\btheta_0,\calA_0) e^{\bfeta_{R10}^{\rm T}\bW} v_{R1,1} \Lambda_{20}(R) = 0
\end{align*}
such that 
\begin{align*} 
e^{\bfeta_{M10}^{\rm T}\bW} v_{M1,1} \Lambda_{10}(t_1) + e^{\bfeta_{R10}^{\rm T}\bW} v_{R1,1} \Lambda_{20}(t_2) = 0
\end{align*}
for any $t_1\in[0,\tau_1]$ and $t_2\in[0,\tau_2]$, leading to $v_{M1,1} = v_{R1,1} = 0$.
Then, equality (\ref{equ:invert_2}) can be reduced to 
\begin{align}
& w_2(\bX;\balpha_0)e^{\bfeta_{M20}^{\rm T}\tbX}\exp\left\{-e^{\bfeta_{M20}^{\rm T}\tbX}\Lambda_{10}(t_1)\right\}e^{ \bfeta_{R20}^{\rm T}\tbX}\exp\left\{-e^{\bfeta_{R20}^{\rm T}\tbX}\Lambda_{20}(t_2)\right\}\nonumber\\
&\qquad\times \left[\tbX^{\rm T}\bv_{M2}\left\{1- e^{\bfeta_{M20}^{\rm T}\tbX}\Lambda_{10}(t_1) \right\} + \tbX^{\rm T}\bv_{R2}\left\{1- e^{\bfeta_{R20}^{\rm T}\tbX} \Lambda_{20}(t_2) \right\} \right] =0.\label{equ:invert_3}
\end{align}

Finally, we set $\Delta^M = \Delta^T=0$ in equality (\ref{equ:infor}) to find 
\begin{align*}
& - \frac{g_3^{(1)}(\btheta_0,\calA_0) e^{\bfeta_{M10}^{\rm T}\bW}\bW^{\rm T}\bv_{M1}+g_3^{(2)}(\btheta_0,\calA_0) e^{\bfeta_{M20}^{\rm T}\tbX}\tbX^{\rm T}\bv_{M2}}{\Psi_3(\btheta_0,\calA_0)} \Lambda_{10}(Y)\\
 &\qquad- \frac{g_2^{(2)}(\btheta_0,\calA_0) e^{\bfeta_{T20}^{\rm T}\tbX}\tbX^{\rm T}\bv_{T2} + g_2^{(3)}(\btheta_0,\calA_0) e^{\bfeta_{T30}^{\rm T}\bW}\bW^{\rm T}\bv_{T3}}{\Psi_3(\btheta_0,\calA_0)} \Lambda_{30}(Y)\\
 &\qquad - \int \frac{g_3^{(1)}(\btheta_0,\calA_0) e^{\bfeta_{M10}^{\rm T}\bW} + g_3^{(2)}(\btheta_0,\calA_0) e^{\bfeta_{M20}^{\rm T}\tbX}}{\Psi_3(\btheta_0,\calA_0)} I(Y\ge t) h_1(t)d\Lambda_{10}(t)\\
&\qquad -\int \frac{g_2^{(2)}(\btheta_0,\calA_0) e^{\bfeta_{T20}^{\rm T}\tbX} + g_2^{(3)}(\btheta_0,\calA_0) e^{\bfeta_{T30}^{\rm T}\bW}}{\Psi_3(\btheta_0,\calA_0)} I(Y\ge t)h_3(t)d\Lambda_{30}(t) = 0,
\end{align*}
such that
\begin{align}
& -g_3^{(2)}(\btheta_0,\calA_0) e^{\bfeta_{M20}^{\rm T}\tbX} \tbX^{\rm T}\bv_{M2} \Lambda_{10}(Y)\nonumber\\
 &\qquad- g_2^{(2)}(\btheta_0,\calA_0) e^{\bfeta_{T20}^{\rm T}\tbX}\tbX^{\rm T}\bv_{T2} \Lambda_{30}(Y) - g_2^{(3)}(\btheta_0,\calA_0) e^{\bfeta_{T30}^{\rm T}\bW}Av_{T3,1} \Lambda_{30}(Y) =0.\label{equ:M0T0}
\end{align}
We set $A=0$ to find 
\begin{align*}
 w_2(\bX;\balpha_0)\exp\left\{-e^{\bfeta_{M20}^{\rm T}\tbX}\Lambda_{10}(t)\right\}e^{\bfeta_{M20}^{\rm T}\tbX} \tbX^{\rm T}\bv_{M2} =0
\end{align*}
for any $t\in[0,\tau_1]$, such that $\bv_{M2}=\bzero$ by Condition \ref{cond:cov}.
Then, equality (\ref{equ:invert_3}) leads to
\begin{align*}
& w_2(\bX;\balpha_0)e^{\bfeta_{M20}^{\rm T}\tbX}\exp\left\{-e^{\bfeta_{M20}^{\rm T}\tbX}\Lambda_{10}(t_1)\right\}e^{ \bfeta_{R20}^{\rm T}\tbX}\exp\left\{-e^{\bfeta_{R20}^{\rm T}\tbX}\Lambda_{20}(t_2)\right\}\\
&\qquad\times \tbX^{\rm T}\bv_{R2}\left\{1- e^{\bfeta_{R20}^{\rm T}\tbX} \Lambda_{20}(t_2) \right\} =0
\end{align*}
for any $t_1\in[0,\tau_1]$ and $t_2\in[0,\tau_2]$, such that $\bv_{R2} = \bzero$.
We then set $A=1$ in equality (\ref{equ:M0T0}) to find
\begin{align*}
&w_2(\bX;\balpha_0) \exp\left\{-e^{\bfeta_{T20}^{\rm T}\tbX}\Lambda_{30}(t)\right\} e^{\bfeta_{T20}^{\rm T}\tbX}\tbX^{\rm T}\bv_{T2} \\
&\qquad+w_3(\bX;\balpha_0) \exp\left\{-e^{\bfeta_{T30}^{\rm T}\tbX}\Lambda_{30}(t)\right\}e^{\bfeta_{T30}^{\rm T}\tbX}v_{T3,1} =0
\end{align*}
for any $t\in[0,\tau_3]$.
Comparing with equality (\ref{equ:invert_1}), we have 
\begin{align*}
& w_2(\bX;\balpha_0)\exp\left\{-e^{\bfeta_{T20}^{\rm T}\tbX}\Lambda_{30}(t)\right\}e^{2\bfeta_{T20}^{\rm T}\tbX}\tbX^{\rm T}\bv_{T2} \Lambda_{30}(t)\\
&\qquad + w_3(\bX;\balpha_0)\exp\left\{-e^{\bfeta_{T30}^{\rm T}\tbX}\Lambda_{30}(t)\right\} e^{2\bfeta_{T30}^{\rm T}\tbX}v_{T3,1} \Lambda_{30}(t) =0
\end{align*}
for any $t\in[0,\tau_3]$.
By Condition \ref{cond:cov} and \ref{cond:alpha}, $v_{T3,1}=0$ and $\bv_{T2}=\bzero$.
We conclude the proof of second identifiability (C7) in \cite{zeng2010general}.
Therefore, Theorem 3 follows from Theorem 2 in \cite{zeng2010general}.
Then, Theorem 4 follows easily by applying functional Delta method.

\bibliography{med_surv}

@article{lin1999nonparametric,
  title={Nonparametric estimation of the gap time distribution for serial events with censored data},
  author={Lin, DY and Sun, W and Ying, Zhiliang},
  journal={Biometrika},
  volume={86},
  number={1},
  pages={59--70},
  year={1999},
  publisher={Oxford University Press}
}

@article{wang1998nonparametric,
  title={Nonparametric estimation of successive duration times under dependent censoring},
  author={Wang, Weijing and Wells, Martin T},
  journal={Biometrika},
  volume={85},
  number={3},
  pages={561--572},
  year={1998},
  publisher={Oxford University Press}
}

@article{vansteelandt2019mediation,
  title={Mediation analysis of time-to-event endpoints accounting for repeatedly measured mediators subject to time-varying confounding},
  author={Vansteelandt, Stijn and Linder, Martin and Vandenberghe, Sjouke and Steen, Johan and Madsen, Jesper},
  journal={Statistics in medicine},
  volume={38},
  number={24},
  pages={4828--4840},
  year={2019},
  publisher={Wiley Online Library}
}

@article{lin2017mediation,
  title={Mediation analysis for a survival outcome with time-varying exposures, mediators, and confounders},
  author={Lin, Sheng-Hsuan and Young, Jessica G and Logan, Roger and VanderWeele, Tyler J},
  journal={Statistics in medicine},
  volume={36},
  number={26},
  pages={4153--4166},
  year={2017},
  publisher={Wiley Online Library}
}

@article{didelez2019defining,
  title={Defining causal mediation with a longitudinal mediator and a survival outcome},
  author={Didelez, Vanessa},
  journal={Lifetime Data Anal.},
  volume={25},
  number={4},
  pages={593--610},
  year={2019},
  publisher={Springer}
}

@article{zeng2010general,
	Author = {Zeng, Donglin and Lin, DY},
	Journal = {Statist. Sin.},
	Number = {2},
	Pages = {871--910},
	Publisher = {NIH Public Access},
	Title = {A general asymptotic theory for maximum likelihood estimation in semiparametric regression models with censored data},
	Volume = {20},
	Year = {2010}}

@article{aalen2020time,
	Author = {Aalen, Odd O and Stensrud, Mats J and Didelez, Vanessa and Daniel, Rhian and R{\o}ysland, Kjetil and Strohmaier, Susanne},
	Journal = {Biometr. J.},
	Number = {3},
	Pages = {532--549},
	Publisher = {Wiley Online Library},
	Title = {Time-dependent mediators in survival analysis: Modeling direct and indirect effects with the additive hazards model},
	Volume = {62},
	Year = {2020}}

@article{mason2015final,
	Author = {Mason, Malcolm D and Parulekar, Wendy R and Sydes, Matthew R and Brundage, Michael and Kirkbride, Peter and Gospodarowicz, Mary and Cowan, Richard and Kostashuk, Edmund C and Anderson, John and Swanson, Gregory and others},
	Journal = {J. Clin. Oncol.},
	Number = {19},
	Pages = {2143--2150},
	Publisher = {American Society of Clinical Oncology},
	Title = {Final report of the intergroup randomized study of combined androgen-deprivation therapy plus radiotherapy versus androgen-deprivation therapy alone in locally advanced prostate cancer},
	Volume = {33},
	Year = {2015}}

@article{angrist1996identification,
	Author = {Angrist, Joshua D and Imbens, Guido W and Rubin, Donald B},
	Journal = {J. Am. Statist. Ass.},
	Number = {434},
	Pages = {444--455},
	Publisher = {Taylor \& Francis},
	Title = {Identification of causal effects using instrumental variables},
	Volume = {91},
	Year = {1996}}

@article{frangakis2002principal,
	Author = {Frangakis, Constantine E and Rubin, Donald B},
	Journal = {Biometrics},
	Number = {1},
	Pages = {21--29},
	Publisher = {Wiley Online Library},
	Title = {Principal stratification in causal inference},
	Volume = {58},
	Year = {2002}}

@article{zhang2003estimation,
	Author = {Zhang, Junni L and Rubin, Donald B},
	Journal = {J. Educ. Behav. Stat.},
	Number = {4},
	Pages = {353--368},
	Publisher = {Sage Publications Sage CA: Los Angeles, CA},
	Title = {Estimation of causal effects via principal stratification when some outcomes are truncated by ``death''},
	Volume = {28},
	Year = {2003}}

@book{klein2006survival,
	Author = {Klein, John P and Moeschberger, Melvin L},
	Publisher = {New York: Springer},
	Title = {Survival Analysis: Techniques for Censored and Truncated Data},
	Year = {2006}}

@article{copelan1991treatment,
	Author = {Copelan, Edward A and Biggs, James C and Thompson, James M and Crilley, Pamela and Szer, Jeff and Klein, John P and Kapoor, Neena and Avalos, Belinda R and Cunningham, Isabel and Atkinson, Kerry},
	Journal = {Blood},
	Number = {3},
	Pages = {838--843},
	Publisher = {Am Soc Hematology},
	Title = {Treatment for Acute Myelocytic Leukemia with Allogeneic Bone Marrow Transplantation Following Preparation with {BuCy2}},
	Volume = {78},
	Year = {1991}}

@article{zheng2017longitudinal,
	Author = {Zheng, Wenjing and van der Laan, Mark},
	Journal = {J. Causal Inference},
	Number = {2},
	Pages = {1--24},
	Publisher = {De Gruyter},
	Title = {Longitudinal Mediation Analysis with Time-varying Mediators and Exposures, with Application to Survival Outcomes},
	Volume = {5},
	Year = {2017}}

@article{lange2012simple,
	Author = {Lange, Theis and Vansteelandt, Stijn and Bekaert, Maarten},
	Journal = {Am. J. Epidemiol.},
	Number = {3},
	Pages = {190--195},
	Publisher = {Oxford University Press},
	Title = {A Simple Unified Approach for Estimating Natural Direct and Indirect Effects},
	Volume = {176},
	Year = {2012}}

@article{xu2010statistical,
	Author = {Xu, Jinfeng and Kalbfleisch, John D and Tai, Beechoo},
	Journal = {Biometrics},
	Number = {3},
	Pages = {716--725},
	Publisher = {Wiley Online Library},
	Title = {Statistical Analysis of Illness--death Processes and Semicompeting Risks Data},
	Volume = {66},
	Year = {2010}}

@article{fine2001semi,
	Author = {Fine, Jason P and Jiang, Hongyu and Chappell, Rick},
	Journal = {Biometrika},
	Number = {4},
	Pages = {907--919},
	Publisher = {Oxford University Press},
	Title = {On Semi-competing Risks Data},
	Volume = {88},
	Year = {2001}}

@article{tchetgen2011causal,
	Author = {Tchetgen Tchetgen, Eric J},
	Journal = {Int. J. Biostat.},
	Number = {1},
	Pages = {1--38},
	Publisher = {De Gruyter},
	Title = {On Causal Mediation Analysis with a Survival Outcome},
	Volume = {7},
	Year = {2011}}

@article{vanderweele2011causal,
	Author = {VanderWeele, Tyler J},
	Journal = {Epidemiol.},
	Number = {4},
	Pages = {582--585},
	Publisher = {NIH Public Access},
	Title = {Causal Mediation Analysis with Survival Data},
	Volume = {22},
	Year = {2011}}

@article{lange2011direct,
	Author = {Lange, Theis and Hansen, J{\o}rgen V},
	Journal = {Epidemiol.},
	Number = {4},
	Pages = {575--581},
	Publisher = {JSTOR},
	Title = {Direct and Indirect Effects in a Survival Context},
	Volume = {22},
	Year = {2011}}

@article{imai2010identification,
	Author = {Imai, Kosuke and Keele, Luke and Yamamoto, Teppei},
	Journal = {Statistical Science},
	Number = {1},
	Pages = {51--71},
	Publisher = {Institute of Mathematical Statistics},
	Title = {Identification, Inference and Sensitivity Analysis for Causal Mediation Effects},
	Volume = {25},
	Year = {2010}}

@article{yu2015semiparametric,
  title={Semiparametric transformation models for causal inference in time-to-event studies with all-or-nothing compliance},
  author={Yu, Wen and Chen, Kani and Sobel, Michael E and Ying, Zhiliang},
  journal={J. R. Statist. Soc. B},
  volume={77},
  number={2},
  pages={397--415},
  year={2015},
  publisher={Wiley Online Library}
}

@article{comment2019survivor,
	Author = {Comment, Leah and Mealli, Fabrizia and Haneuse, Sebastien and Zigler, Corwin},
	Journal = {arXiv preprint arXiv:1902.09304},
	Title = {Survivor average causal effects for continuous time: a principal stratification approach to causal inference with semicompeting risks},
	Year = {2019}}

@article{maller1992estimating,
  title={Estimating the proportion of immunes in a censored sample},
  author={Maller, Ross A and Zhou, S},
  journal={Biometrika},
  volume={79},
  number={4},
  pages={731--739},
  year={1992},
  publisher={Oxford University Press}
}

@article{huang2020causal,
  title={Causal mediation of semicompeting risks
},
  author={Huang, Yen-Tsung},
  journal={Biometrics},
  year={2020}
}
\end{document}